\RequirePackage{lineno}
\documentclass[aps,prd,twocolumn,showpacs,superscriptaddress,groupedaddress]{revtex4}  
\usepackage{graphicx}  
\usepackage{dcolumn}   
\usepackage{bm}        
\usepackage{amssymb}   
\usepackage{longtable}
\usepackage{epsfig}

\def\ppbar{{p}\overline{{p}}}

\def\METnoSpace{{\mbox{$E\kern-0.57em\raise0.19ex\hbox{/}_{T}$}}}

\def\met{\ifmmode
        {\hbox{E\kern-0.6em\lower-.1ex\hbox{/}}_T }          
        \else
        {${\hbox{$E$\kern-0.6em\lower-.1ex\hbox{/}}}_T$}\fi}

\providecommand{\e}[1]{\ensuremath{\times 10^{#1}}}

\begin{document}

\hspace{5.2in} \mbox{Fermilab-Pub-13-103-E}

\title{Measurement of the $\boldsymbol{ZZ}$ production cross section and 
search for the standard model Higgs boson in the four lepton final state 
in $\boldsymbol{p \bar{p}}$ collisions}
\affiliation{LAFEX, Centro Brasileiro de Pesquisas F\'{i}sicas, Rio de Janeiro, Brazil}
\affiliation{Universidade do Estado do Rio de Janeiro, Rio de Janeiro, Brazil}
\affiliation{Universidade Federal do ABC, Santo Andr\'e, Brazil}
\affiliation{University of Science and Technology of China, Hefei, People's Republic of China}
\affiliation{Universidad de los Andes, Bogot\'a, Colombia}
\affiliation{Charles University, Faculty of Mathematics and Physics, Center for Particle Physics, Prague, Czech Republic}
\affiliation{Czech Technical University in Prague, Prague, Czech Republic}
\affiliation{Institute of Physics, Academy of Sciences of the Czech Republic, Prague, Czech Republic}
\affiliation{Universidad San Francisco de Quito, Quito, Ecuador}
\affiliation{LPC, Universit\'e Blaise Pascal, CNRS/IN2P3, Clermont, France}
\affiliation{LPSC, Universit\'e Joseph Fourier Grenoble 1, CNRS/IN2P3, Institut National Polytechnique de Grenoble, Grenoble, France}
\affiliation{CPPM, Aix-Marseille Universit\'e, CNRS/IN2P3, Marseille, France}
\affiliation{LAL, Universit\'e Paris-Sud, CNRS/IN2P3, Orsay, France}
\affiliation{LPNHE, Universit\'es Paris VI and VII, CNRS/IN2P3, Paris, France}
\affiliation{CEA, Irfu, SPP, Saclay, France}
\affiliation{IPHC, Universit\'e de Strasbourg, CNRS/IN2P3, Strasbourg, France}
\affiliation{IPNL, Universit\'e Lyon 1, CNRS/IN2P3, Villeurbanne, France and Universit\'e de Lyon, Lyon, France}
\affiliation{III. Physikalisches Institut A, RWTH Aachen University, Aachen, Germany}
\affiliation{Physikalisches Institut, Universit\"at Freiburg, Freiburg, Germany}
\affiliation{II. Physikalisches Institut, Georg-August-Universit\"at G\"ottingen, G\"ottingen, Germany}
\affiliation{Institut f\"ur Physik, Universit\"at Mainz, Mainz, Germany}
\affiliation{Ludwig-Maximilians-Universit\"at M\"unchen, M\"unchen, Germany}
\affiliation{Panjab University, Chandigarh, India}
\affiliation{Delhi University, Delhi, India}
\affiliation{Tata Institute of Fundamental Research, Mumbai, India}
\affiliation{University College Dublin, Dublin, Ireland}
\affiliation{Korea Detector Laboratory, Korea University, Seoul, Korea}
\affiliation{CINVESTAV, Mexico City, Mexico}
\affiliation{Nikhef, Science Park, Amsterdam, the Netherlands}
\affiliation{Radboud University Nijmegen, Nijmegen, the Netherlands}
\affiliation{Joint Institute for Nuclear Research, Dubna, Russia}
\affiliation{Institute for Theoretical and Experimental Physics, Moscow, Russia}
\affiliation{Moscow State University, Moscow, Russia}
\affiliation{Institute for High Energy Physics, Protvino, Russia}
\affiliation{Petersburg Nuclear Physics Institute, St. Petersburg, Russia}
\affiliation{Instituci\'{o} Catalana de Recerca i Estudis Avan\c{c}ats (ICREA) and Institut de F\'{i}sica d'Altes Energies (IFAE), Barcelona, Spain}
\affiliation{Uppsala University, Uppsala, Sweden}
\affiliation{Lancaster University, Lancaster LA1 4YB, United Kingdom}
\affiliation{Imperial College London, London SW7 2AZ, United Kingdom}
\affiliation{The University of Manchester, Manchester M13 9PL, United Kingdom}
\affiliation{University of Arizona, Tucson, Arizona 85721, USA}
\affiliation{University of California Riverside, Riverside, California 92521, USA}
\affiliation{Florida State University, Tallahassee, Florida 32306, USA}
\affiliation{Fermi National Accelerator Laboratory, Batavia, Illinois 60510, USA}
\affiliation{University of Illinois at Chicago, Chicago, Illinois 60607, USA}
\affiliation{Northern Illinois University, DeKalb, Illinois 60115, USA}
\affiliation{Northwestern University, Evanston, Illinois 60208, USA}
\affiliation{Indiana University, Bloomington, Indiana 47405, USA}
\affiliation{Purdue University Calumet, Hammond, Indiana 46323, USA}
\affiliation{University of Notre Dame, Notre Dame, Indiana 46556, USA}
\affiliation{Iowa State University, Ames, Iowa 50011, USA}
\affiliation{University of Kansas, Lawrence, Kansas 66045, USA}
\affiliation{Louisiana Tech University, Ruston, Louisiana 71272, USA}
\affiliation{Northeastern University, Boston, Massachusetts 02115, USA}
\affiliation{University of Michigan, Ann Arbor, Michigan 48109, USA}
\affiliation{Michigan State University, East Lansing, Michigan 48824, USA}
\affiliation{University of Mississippi, University, Mississippi 38677, USA}
\affiliation{University of Nebraska, Lincoln, Nebraska 68588, USA}
\affiliation{Rutgers University, Piscataway, New Jersey 08855, USA}
\affiliation{Princeton University, Princeton, New Jersey 08544, USA}
\affiliation{State University of New York, Buffalo, New York 14260, USA}
\affiliation{University of Rochester, Rochester, New York 14627, USA}
\affiliation{State University of New York, Stony Brook, New York 11794, USA}
\affiliation{Brookhaven National Laboratory, Upton, New York 11973, USA}
\affiliation{Langston University, Langston, Oklahoma 73050, USA}
\affiliation{University of Oklahoma, Norman, Oklahoma 73019, USA}
\affiliation{Oklahoma State University, Stillwater, Oklahoma 74078, USA}
\affiliation{Brown University, Providence, Rhode Island 02912, USA}
\affiliation{University of Texas, Arlington, Texas 76019, USA}
\affiliation{Southern Methodist University, Dallas, Texas 75275, USA}
\affiliation{Rice University, Houston, Texas 77005, USA}
\affiliation{University of Virginia, Charlottesville, Virginia 22904, USA}
\affiliation{University of Washington, Seattle, Washington 98195, USA}
\author{V.M.~Abazov} \affiliation{Joint Institute for Nuclear Research, Dubna, Russia}
\author{B.~Abbott} \affiliation{University of Oklahoma, Norman, Oklahoma 73019, USA}
\author{B.S.~Acharya} \affiliation{Tata Institute of Fundamental Research, Mumbai, India}
\author{M.~Adams} \affiliation{University of Illinois at Chicago, Chicago, Illinois 60607, USA}
\author{T.~Adams} \affiliation{Florida State University, Tallahassee, Florida 32306, USA}
\author{J.P.~Agnew} \affiliation{The University of Manchester, Manchester M13 9PL, United Kingdom}
\author{G.D.~Alexeev} \affiliation{Joint Institute for Nuclear Research, Dubna, Russia}
\author{G.~Alkhazov} \affiliation{Petersburg Nuclear Physics Institute, St. Petersburg, Russia}
\author{A.~Alton$^{a}$} \affiliation{University of Michigan, Ann Arbor, Michigan 48109, USA}
\author{A.~Askew} \affiliation{Florida State University, Tallahassee, Florida 32306, USA}
% ...
%
%
\collaboration{The D0 Collaboration\footnote{with visitors from
$^{a}$Augustana College, Sioux Falls, SD, USA,
$^{b}$The University of Liverpool, Liverpool, UK,
$^{c}$DESY, Hamburg, Germany,
$^{d}$Universidad Michoacana de San Nicolas de Hidalgo, Morelia, Mexico
$^{e}$SLAC, Menlo Park, CA, USA,
$^{f}$University College London, London, UK,
$^{g}$Centro de Investigacion en Computacion - IPN, Mexico City, Mexico,
$^{h}$Universidade Estadual Paulista, S\~ao Paulo, Brazil,
$^{i}$Karlsruher Institut f\"ur Technologie (KIT) - Steinbuch Centre for Computing (SCC)
and
$^{j}$Office of Science, U.S. Department of Energy, Washington, D.C. 20585, USA.
$^{k}$Visitor from Lewis University, Romeoville, IL, USA.
}} \noaffiliation
\vskip 0.25cm
\date{April 19, 2013}

\begin{abstract}
We present a measurement of $Z$ boson pair production in $p \bar{p}$ 
collisions at 1.96 TeV with 9.6~fb$^{-1}$ to 
9.8~fb$^{-1}$ of D0 data.
We examine the final states  
$eeee$, $ee\mu\mu$, and $\mu\mu\mu\mu$.
Based on selected data, the measured cross section
in the mass region $M(Z/\gamma^{*}) > 30$~GeV is
$\sigma(p\bar{p} \to Z/\gamma^*\mbox{ }Z/\gamma^*)= 1.26 ^{+0.44}_{-0.36} \thinspace \mbox{(stat)}  ^{+0.17}_{-0.15} \thinspace \mbox{(syst)} \pm 0.08 \thinspace \mbox{(lumi)}$~pb; after correcting 
for the expected ratio of $\sigma(p\bar{p} \to Z/\gamma^*\mbox{ }Z/\gamma^*)$ to $\sigma(p\bar{p} \to ZZ)$, we derive a cross section for 
$p\bar{p} \to ZZ$ production of $1.05 ^{+0.37}_{-0.30} \thinspace \mbox{(stat)}  ^{+0.14}_{-0.12} \thinspace \mbox{(syst)} \pm 0.06 \thinspace \mbox{(lumi)}$~pb. 
This result is combined with a previous result from the $ZZ\to\ell^{+}\ell^{-}\nu\bar{\nu}$ channel resulting in a combined $p\bar{p} \to ZZ$ 
cross section measurement of 
$1.32 ^{+0.29}_{-0.25} \thinspace \mbox{(stat)}  \pm 0.12 \thinspace \mbox{(syst)} \pm 0.04\thinspace \mbox{(lumi)}$~pb.
These measurements are
consistent with the standard model expectation of $1.43 \pm 0.10$ pb.  We extend this analysis to search for the standard model 
(SM) Higgs boson between 115 and 200 GeV.
At a Higgs boson mass of 125 GeV, we expect to set a limit of 43 times the SM expectation at 95\% C.L., and 
set a limit of 42 times the SM expectation at 95\% C.L.
\end{abstract}


\maketitle

\newpage

\section{\label{sec:introduction}Introduction}
We present a measurement of the cross section $\sigma(p\bar{p} \to Z/\gamma^{*}\mbox{ }Z/\gamma^{*})$
at $\sqrt{s} = 1.96$~TeV, using events where each $Z/\gamma^{*}$ 
results in two charged leptons.  
Because the branching fraction of the $Z$ boson to charged leptons is smaller than that to quarks or neutrinos, this 
process is relatively rare, but has the advantage of being an extremely pure final state.  The largest fraction of the 
background results from events in which one or more jet has been misidentified as a lepton,
since few other processes in the standard model (SM) produce four isolated leptons.  We also unfold our measurement to 
determine the $\sigma(p\bar{p} \to ZZ)$ cross section.

After measuring the $t$-channel $Z/\gamma^{*}\mbox{ }Z/\gamma^{*}$ cross section, we reinterpret the analysis 
as a search for the Higgs boson in the four lepton final state, predicted in the SM as a result of electroweak symmetry breaking.  
Both the ATLAS and CMS experiments at the CERN LHC $pp$ collider have observed a four lepton resonance at 
a mass of $\sim$125 GeV \cite{atlas_higgs, cms_higgs} which, when combined with other decay channels, is consistent with the SM Higgs boson.  

$Z$ boson pair production was studied at the CERN LEP2 collider by the ALEPH \cite{Barate:1999jj}, DELPHI
 \cite{Abdallah:2003dv}, L3 \cite{Acciarri:1999ug}, and  
OPAL \cite{Abbiendi:2003va} collaborations in multiple final states, including 
$e^+e^- \rightarrow \ell^{+} \ell^{-} \ell^{'+} \ell^{'-}$, where $\ell$ represents an electron or a muon.  The LEP experiments also
set limits on anomalous $ZZZ$  and $ZZ\gamma$ couplings \cite{Alcaraz:2006mx}.

The Fermilab Tevatron experiments have also searched for and measured the pair
production of $Z$ bosons.  The D0 collaboration's analysis of 
$ZZ \rightarrow \ell^+ \ell^- \ell^{'+} \ell^{'-}$ production with 1.1 fb$^{-1}$ of $p \bar{p}$ data yielded an upper limit of
4.4 pb on the $ZZ$ production cross section at 95\% C.L. Additionally,
limits on anomalous $ZZZ$ and $ZZ\gamma$ couplings were 
determined~\cite{:2007hm}.  The D0 collaboration was the first to observe $ZZ$ production in $p \bar{p}$ 
collisions in the $\ell^+ \ell^- \ell^{'+} \ell^{'-}$ final state with 
2.7 fb$^{-1}$ of data \cite{iananalysis}. 
The D0 collaboration has also measured the $ZZ$ cross section in the $\ell^+\ell^- \nu \bar{\nu}$ final state, 
first with 2.2 fb$^{-1}$ \cite{d0_zz_2l2nu1} and later with
8.6 fb$^{-1}$ of integrated luminosity, yielding a final measurement 
of $1.64 \pm 0.44 \thinspace \mbox{(stat)}^{+0.13}_{-0.15}\thinspace \mbox{(syst)}$ pb~\cite{d0_zz_2l2nu}.
The CDF collaboration has analyzed data from 1.9~fb$^{-1}$ of integrated luminosity to study $ZZ$ 
production, measuring, when combining  
$\ell^+ \ell^- \ell^{'+} \ell^{'-}$  and  $\ell^+\ell^- \nu \bar{\nu}$ 
channels, a cross section of
$\sigma(ZZ) = 1.4^{+0.7}_{-0.6}~\mathrm{(stat+syst)}$~pb~\cite{cdf_zz}.  
The ATLAS collaboration has observed  
$pp \to ZZ$ production in the four charged lepton final state 
in 1.0 fb$^{-1}$ of data at $\sqrt{s} = 7$ TeV \cite{atlas_zz}.
The CMS collaboration has measured $\sigma(pp \to ZZ)$ 
in 5.0 fb$^{-1}$ of data at $\sqrt{s} = 7$ TeV \cite{cms_zz}, and 
has observed the rare decay $Z \to \ell^+ \ell^- \ell^{'+} \ell^{'-}$ with a branching fraction in agreement with the SM prediction.

This article is an update of the D0 collaboration's prior $ZZ$ to four charged lepton analysis that 
measured a cross section of $\sigma(p\bar{p} \to
ZZ)=1.26^{+0.47}_{-0.37}~\mathrm{(stat)} \pm 0.11~\mathrm{(syst)} \pm
0.08~\mathrm{(lumi)}$~pb  using 6.4~fb$^{-1}$ of integrated luminosity \cite{d0_zz_4lep}.  
The result presented here uses 9.6~fb$^{-1}$ to 9.8~fb$^{-1}$ of integrated luminosity, and expands electron acceptance 
in the $eeee$ final state.

\section{\label{sec:detector}Detector}

The D0 detector is described in detail elsewhere~\cite{thedetector,thedetector2,thedetector3,thedetector4}.
The main components are the central tracking system, the
calorimeter system, and the muon detectors. The central-tracking system is
located within a 2~T solenoidal  field and consists of two different trackers.
Located closest to the interaction point is the silicon microstrip tracker (SMT)
and surrounding that is the  central 
fiber tracker (CFT).  The SMT is an assembly of 
barrel silicon detectors in the central region, along with large-diameter disks in
the forward regions for tracking at  high pseudorapidity ($\eta$) \cite{etaref}.  
The CFT consists of eight concentric coaxial barrels each carrying two doublet layers of 
scintillating fibers.  
The  liquid-argon calorimeter system 
is housed in three cryostats.   The central calorimeter (CC)  covers up to $|\eta|=1$, and two end 
calorimeters (EC) 
are located in the forward regions, extending coverage to $|\eta|=4$.  In the 
intercryostat region (ICR) between the CC and EC cryostats, there is a scintillating
intercryostat detector (ICD) between $1.1 <|\eta| <1.4$ that recovers some energy from particles 
passing through the ICR.
Closest to the collisions are the
electromagnetic (EM) regions of the  calorimeter followed by hadronic layers of
fine and coarse segmentation.  

A muon detection system \cite{themuons1} is located beyond the calorimeters and consists of a layer of tracking detectors and 
scintillation trigger counters before 1.8 T toroid magnets, followed by two similar layers after the toroids.

There is a three-level trigger system consisting of a
collection of specialized  hardware elements, microprocessors, and decision-making algorithms 
to selectively record the events of most interest.

\section{Monte Carlo}

\begin{figure}[!htb]
\begin{minipage}[t]{0.45\linewidth}
\centering
\includegraphics[width=1.5in]{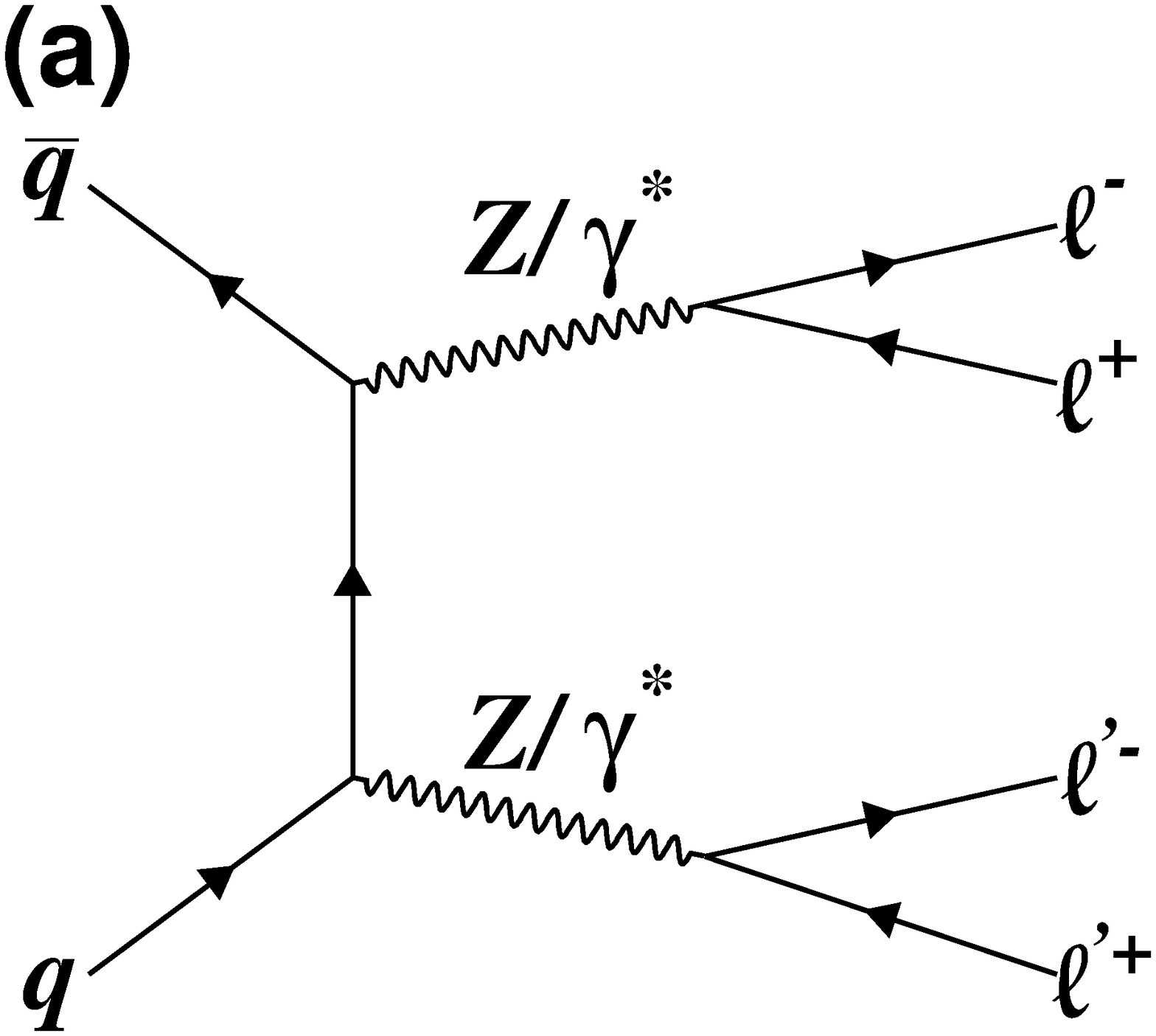}
 \label{fig:decay}
\end{minipage}
\begin{minipage}[t]{0.45\linewidth}
\centering
\includegraphics[width=1.5in]{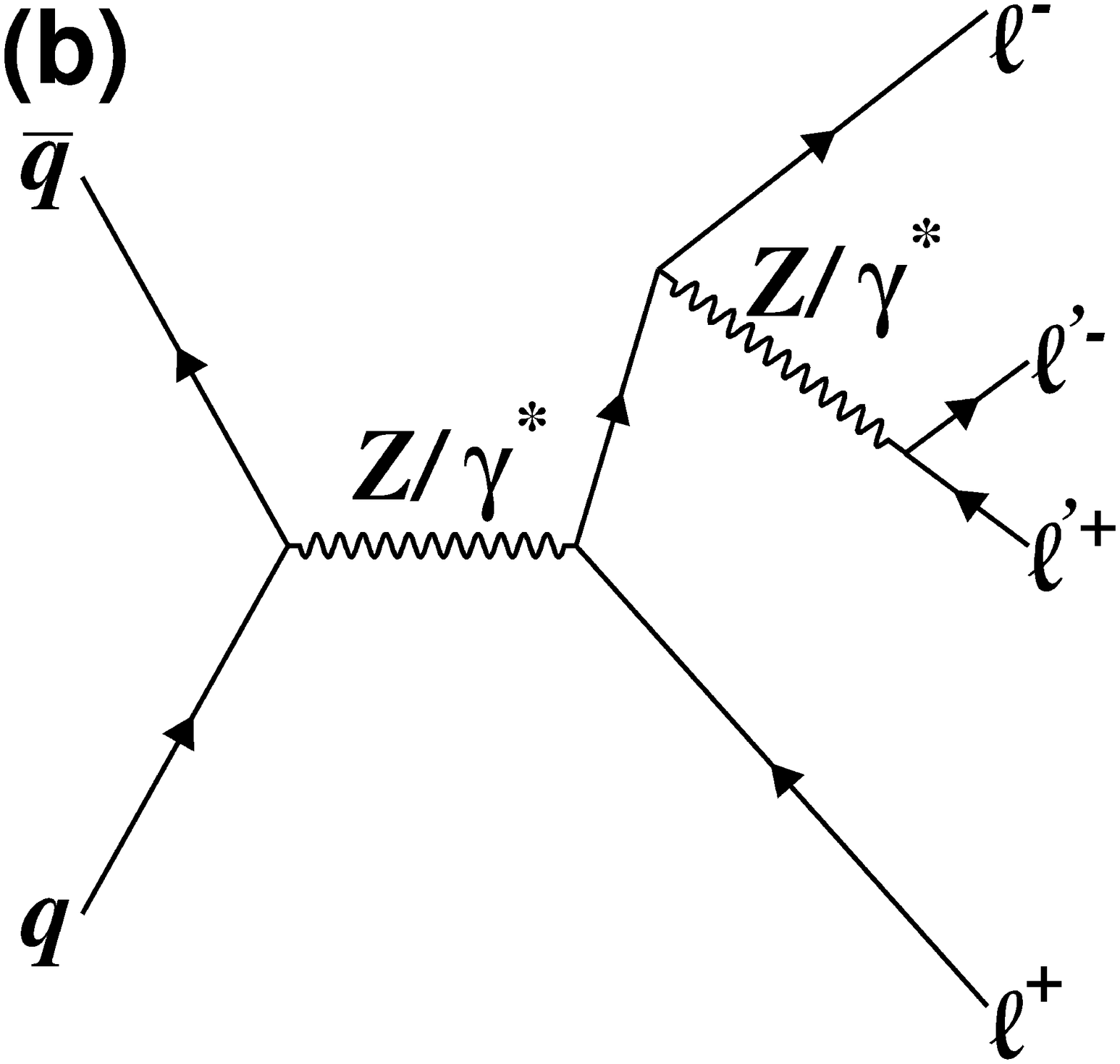}
\label{fig:singledecay}
\end{minipage}
\caption{Feynman diagrams for (a) the $t$-channel tree-level process  $q\bar{q} \rightarrow ZZ \to \ell^+ \ell^- \ell^{'+} \ell^{'-}$ and (b) the 
singly resonant process.}
\label{fig:decayfd}
\end{figure}

We use the {\sc pythia} ~\cite{Pythia} Monte Carlo (MC) program to determine the 
$Z/\gamma^{*} \mbox{ } Z/\gamma^{*} \to \ell^+ \ell^- \ell^{'+} \ell^{'-}$ signal acceptance and to simulate the migration background.  
The signal is defined to consist of $Z/\gamma^{*} \mbox{ }Z/\gamma^{*}$ pairs where each $Z/\gamma^{*}$ boson has a 
mass greater than 30 GeV.  The migration background consists of $Z/\gamma^{*}\mbox{ } Z/\gamma^{*}$ 
events where at least one of the two  $Z/\gamma^{*}$ bosons has an invariant mass of less than
30 GeV; it enters the signal sample either due to mismeasurement or by mis-assigning 
the lepton pairs in the $eeee$ and $\mu \mu \mu \mu$ channels.  
We include  
$Z/\gamma^{*} \mbox{ } Z/\gamma^{*} \to \ell^+ \ell^- \tau^{+} \tau^{-}$ events where the taus decay into
electrons or muons as appropriate to match the final four-lepton signature in the signal 
acceptance.
Contributions from $ZZ \to \tau^{+} \tau^{-} \tau^{+} \tau^{-}$ with subsequent decays into
muons and electrons are  also examined, but found to be negligible.
The $ZZ$ transverse momentum ($p_T$) spectrum is also estimated using {\sc sherpa} MC \cite{sherpa}, and the difference 
between the $p_T$ spectra from {\sc pythia} and {\sc sherpa} is used as a systematic.   
The dominant tree-level diagrams for $p\bar{p} \rightarrow Z/\gamma^* \mbox{ }Z/\gamma^* \rightarrow \ell^+ \ell^- \ell^{'+} \ell^{'-}$ are shown 
in Fig.~\ref{fig:decayfd}.  The singly resonant $Z$ boson diagram contributes at low mass, and 
we expect a negligible contribution to the signal yields from this diagram 
in our analysis.  

To estimate the migration background, we generate $Z/\gamma^{*}$ pairs where at least one of the bosons 
has a mass between 5 and 30 GeV, and estimate the cross section of these events 
using next-to-leading-order (NLO) MC from {\sc mcfm}~\cite{MCFM} with the CTEQ61M PDF set \cite{cteq6m}. 

The $t \bar t$ background is estimated using {\sc alpgen}~\cite{Alpgen} with a top quark
mass of 172~GeV and is normalized to an approximate NNLO cross section calculation~\cite{Moch}. 

Other backgrounds where photons or jets must be misidentified for the event to enter our sample,
such as events containing a $Z$ plus jets, are estimated from data as described in Sec.~\ref{sec:InstrumBkg}.

For the Higgs boson search, we generate SM Higgs boson events with masses between 115 and 200 GeV in 5 GeV 
increments.  We simulate the gluon fusion ($gg \to H$) and $ZH$ associated production ($q\bar{q} \to ZH$) processes
using {\sc pythia}.  The expected $gg \to H$ 
cross section is corrected to next-to-NLO (NNLO) with next-to-next-to-leading-log resummation of soft gluons ~\cite{higgsxsec1}.  
The associated $ZH$ production cross section is corrected to NNLO \cite{higgsxec2}.
The expected branching fractions for the Higgs boson decay are determined using {\sc hdecay} \cite{hdecay}.  

All of the MC samples are passed through a {\sc geant} \cite{geant} simulation of the D0 detector.
To account for detector noise and additional $p \bar{p}$ interactions, data from random beam 
crossings are overlaid onto all MC events to match the instantaneous luminosity distribution 
of the selected data.  The same algorithms used to reconstruct real data events 
are run on these simulated events.

\section{Object Identification \label{sec:objid}}

All muon candidates are reconstructed either as a muon track reconstructed from hits in 
both the wire chambers and 
scintillators in at least one layer of the muon system, or as a narrow energy deposit in the calorimeter system consistent with that 
expected from a muon 
passing through the calorimeter that is not associated with tracks in the muon system.  Each muon candidate must be  
matched to a track in 
the central tracker with a $p_T$ $>$ 15 GeV, and the track $p_T$ is taken as the $p_T$ of the 
muon, $p_T^{\mu}$.  This track must have an impact parameter 
consistent with the muon coming from the interaction point.  We consider two muon isolation variables:  $E_{T}^{\rm trkcone}$,
the scalar sum of the track $p_T$ 
within a cone of $\Delta R \leq 0.5$~\cite{delRdef} about the muon track; and $E_{T}^{\rm halo}$, the sum of the calorimeter energy in an annulus $ 0.1 <\Delta R  \leq 0.4$ centered on the muon track.  If the muon is reconstructed in the muon system, then 
we impose the requirement that  $E_{T}^{\rm trkcone} / p_T^{\mu} < 0.25$ and $E_{T}^{\rm halo}/p_T^{\mu}<0.4$.  
Otherwise, each variable divided by $p_T^{\mu}$ must be less than $0.1$.

Different selection requirements apply for electrons identified in the CC ($|\eta_{d}|<1.1$), EC ($1.5 < | \eta_{d}| < 3.2$), and 
ICR ($1.1 < | \eta_{d}| < 1.5$), where $\eta_{d}$ is the pseudorapidity calculated with respect to the center 
of the detector.  In the CC and EC, electrons must have at least 90\% of their energy found in the EM calorimeter, have $p_T>$ 15 GeV,
and pass a calorimeter isolation requirement.  The $p_T$ estimate for the CC and EC electrons is based on the energy deposited in the 
calorimeter.  
For electrons in the CC, the sum of transverse momenta of the charged central tracks in an annulus of 
$ 0.05 <\Delta R  \leq 0.4$ about the electron, $I4$, must be less than 4.0 GeV.  There must either be a track in the central 
tracker associated with the calorimeter cluster, or hits in the central tracker consistent with a track along the 
extrapolation of the calorimeter 
cluster to the interaction point.  Finally, the electron must pass a neural net (NN) discriminant trained to separate electrons from jets in the CC using seven shower shape and isolation variables as input.

In the EC only, we require that the 
track isolation $I4$ be less than  
$(7.0 -2.5 \times |\eta_{d}|)$ GeV or $0.01$ GeV, whichever is larger.  
The electron must pass a NN discriminant trained to separate electrons from jets in the EC using three shower shape and isolation variables as input and an additional chi-square-based shower shape requirement designed to distinguish electrons from jets.

Within the ICR, there is incomplete EM calorimeter coverage, so the electron must pass a minimum EM + ICD energy fraction requirement 
that varies with $|\eta_{d}|$.  
The candidate must be matched to a central track 
with $p_T > 15$ GeV and have a $p_T > 10$ GeV measured in the calorimeter.  Additionally, the ICR electron must satisfy 
two multivariate discriminants designed to reject jet background.  Due to the limited energy resolution in the ICR, 
we use the $p_T$ of the track associated with the ICR electron to estimate the ICR electron energy.  

Jets are used in the estimation of the instrumental background, as discussed in Sec.~\ref{sec:InstrumBkg}.  In this analysis, we use jets reconstructed 
from energy deposits in the CC, EC, and ICD detectors using the Run II midpoint cone algorithm \cite{jccbjets} 
with a cone size of $\Delta R = 0.5$.  The jets must have $p_T > 15$ GeV and $|\eta_{d}| < 3.2$.  
We apply the standard jet energy scale (JES) corrections \cite{jes} to jets in both data and MC.

The missing transverse energy, \met, is calculated using a
vector sum of the transverse components of calorimeter energy depositions, with 
appropriate JES corrections \cite{jes}.  In the $ee\mu \mu$ and $\mu\mu\mu\mu$ final states, the \met~is 
corrected for identified muons.  

\section{\label{sec:selection} Event Selection}

To maximize the acceptance, we consider all events that pass the event selection requirements below
without requiring a specific trigger.  The majority of our acceptance comes from events collected by 
single lepton and di-lepton triggers.  As there are four high-$p_{T}$  leptons in this final state, 
we estimate that the trigger efficiency for the signal is greater than 99.5\% in all channels.  

\subsection{\label{sec:electrons}$\boldsymbol{ eeee}$ final state}
All electron candidates have to satisfy the requirements in Sec.~\ref{sec:objid}.
We require at least four electron candidates.  
If there are four CC/EC electron candidates, no ICR electron candidates are considered, and 
if there are more than four CC/EC electron candidates, the highest-$p_T$ candidates are used.  
At least two of the electrons must be in the CC, and if an event has more than one ICR electron, 
only the leading ICR electron is considered as a lepton candidate.
All possible pairings of the selected electrons are 
considered with no charge requirement imposed, and we require that one of the pairings has di-electron mass 
$M_{ee}>30$ GeV for both di-electrons.  Additionally, there must be $\Delta R  > 0.5$ between any ICR electron and any CC and EC electrons, or the ICR electron is not considered.  
Because the instrumental background contamination is expected to vary significantly depending on the number of
central electrons, the $eeee$ channel is then divided into four sub-channels that depend on
the number of electrons in the CC, $N_{CC}$, the EC, $N_{EC}$, and in the ICR, $N_{ICR}$:   $N_{CC}=4$, $N_{CC}=2$ 
with $N_{EC}=2$, $N_{CC}=3$ with $N_{EC}=1$, and $N_{CC} \ge 2$ with $N_{ICR}=1$.
Since we do not use the muon system in $eeee$ event reconstruction, we include 
events where the muon system was not fully operational.  This leads to a slightly higher integrated luminosity in the $eeee$ final state 
compared to the $ee \mu \mu$ and $\mu \mu \mu \mu$ final states. 

\subsection{\label{sec:emu}$\boldsymbol{ ee\mu\mu}$ channel}

The $ee\mu\mu$ channel is divided into three sub-channels that
depend on the number of electrons in the CC: $N_{CC}=2$,
$N_{CC}=1$, and $N_{CC} = 0$.   No ICR electrons are used in this channel.
As in the $eeee$ final state, we apply this splitting because the instrumental background
contamination varies significantly depending on  
the number of central electrons.  We require at least two electrons and two muons;  if 
there are more leptons in the event, only the highest-$p_T$ leptons of each type are used.  
To reject cosmic ray background, the cosine of the angle between the muons must satisfy $\mbox{cos}\alpha <0.96$, and the 
acoplanarity \cite{cit:acoplanarity} between the two muons must be greater than 0.05 radians.  We further require $|\Delta z_{DCA}| < 3.0$ cm between 
the muon tracks, where $z_{DCA}$ refers to the location along the beam axis where the track has its distance of closest approach to 
the beamline.  Also, we impose the requirement that $\Delta R > 0.2$ between all possible electron-muon 
pairings.  Both the muon pair and electron pair invariant masses must exceed 30 GeV.  There is no opposite charge requirement placed on 
the lepton pairs in order to maximize acceptance.

\subsection{\label{sec:muons}$\boldsymbol{\mu\mu\mu\mu}$ final state}
In the four-muon final state, there must be at least four muon candidates satisfying the requirements in Sec.~\ref{sec:objid}, 
and at least two of the muons must be matched to tracks found in the muon system.  
The four-muon system must be charge neutral ($\sum_{i=1}^{4}q_{i} = 0$), and only oppositely charged pairs are considered as $Z$ boson 
candidates.  If more than four muons are reconstructed in the event, we consider only the four highest-$p_T$ muons.  
We further require $|\Delta z_{DCA}| < 3.0$ cm between all muons.
We also require that one of the two possible sets of dimuons has a dimuon mass $M_{\mu\mu}>30$ GeV  for both dimuons.

\section{Instrumental Background \label{sec:InstrumBkg}}
The instrumental background primarily arises from $Z(\to \ell \ell)$ + jets and $Z(\to \ell \ell) + \gamma$ +
jets production (with smaller contributions from $WZ$ + jets, $WW$ + jets, $W$ + jets, and multijet production with $\ge$4 jets).
These events contaminate the four-lepton channels when a jet is falsely reconstructed as an isolated lepton.  
$Z(\to \ell \ell) + \gamma$ + jets
production where a photon and a jet are mis-identified as an electron contaminates the $eeee$ and $ee\mu\mu$ channels.  

We estimate the instrumental background using the data.  We first find the 
probability for a jet to be mis-identified as a lepton, $P_{j\ell}$. 
A tag and probe method is used to determine $P_{j\ell}$ 
where di-jet activity is considered with jet $p_T > 15$~GeV. 
The tagged jet must be associated with a jet that fired a single jet trigger and be the 
highest-$p_T$ jet in the event.  We then look for a probe jet with 
$|\Delta \phi| > 3.0$ with respect to the tag jet, where $\phi$ is the azimuthal angle.  To suppress contamination
from $W$+jet events, we require  $\mbox{\met}< 20$ GeV in the tag and probe sample.  The probe jets form 
the denominator of the $P_{j\ell}$ calculation.

To calculate the numerator of the $P_{je}$ estimate, we first find all good electrons in the event 
with a $p_T > 15$ GeV.  We then select those electrons that satisfy the same criteria imposed on the  
probe jets, noted above. The $P_{je}$ estimate is parametrized as a function of the jet $p_T$ and $\eta_{d}$.

The $P_{j\mu}$ estimate is determined using a similar method.  The tagged jet is defined as 
was done for electron events, but in the numerator, rather than have an electron, 
we use any muon that has $|\Delta \phi| > 3.0$ from the tag jet, and $P_{j\mu}$ 
is taken as the number of muons divided by the number of probe jets in the sample.  The 
$P_{j\mu}$ estimate is parameterized in terms of $p_{T}$ and $\eta$.

The $P_{j\ell}$ estimates for both electrons and muons are on the order of 10$^{-3}$.

To estimate the instrumental background for the $eeee$ final state, $P_{je}$ is applied to events with 
three reconstructed electrons and one or more jet.  The jet kinematics are used to model 
the electron kinematics in the event.  This method accounts for events where either 
a photon or a jet is misreconstructed as one electron and a jet is misreconstructed as the other.  This method overestimates 
the background from events with two real electrons and two jets misreconstructed as electrons.  To determine the rate, 
we look at events with two reconstructed electrons and two or more reconstructed jets and 
apply $P_{je}$ to both jets.  The number of $ee$ plus two jet events after $P_{je}$ is applied to both jets is 
found to be negligible, so only $eee$+jet events are used to model the instrumental background distributions in 
the $eeee$ final state.

The instrumental background in the  $ee\mu\mu$ channel is calculated from two different 
contributions. The first contribution is from events with $e \mu \mu$ plus one or more jet, where we apply $P_{je}$
to the jet.  This  method gives an estimate of a background due to
$Z(\to\mu\mu)$ + jets and $Z(\to\mu\mu)+ \gamma$ + jets where a jet has been reconstructed as an
electron.  We also consider the $ee$ plus two jet or more case, where we apply $P_{j\mu}$ to the jets. This method gives an
estimate of the  background due to $Z(\to ee)$ + jets where the jets can contain muons.

The $P_{j\mu}$ is applied to jets in $\mu\mu$ plus two or more jets data to
determine the instrumental background for the $\mu\mu\mu\mu$ channel.

Background estimates derived from the above method can be found in Tables~\ref{tab:embkgd}--\ref{tab:mubkgd} in each final state.

\begin{table*}[htb]
\caption{\label{tab:embkgd}Contributions from non-negligible
backgrounds in the $eeee$ subchannels, plus expected $t$-channel $ZZ$ and Higgs boson signals and
number of observed events. Uncertainties are statistical followed by
systematic.}
\begin{tabular}{l|c|c|c|c}
\hline \hline
           & 2 CC & 3 CC & 4 CC & $\geq$ 2 CC \\
           & 2 EC & 1 EC &      & 1 ICR \\
    \hline
& & & & \\[-2mm]
    Instrumental backg.
    & $0.15 \pm 0.01 \pm 0.03$
    & $0.12 \pm 0.01 \pm 0.02$
    & $0.05 \pm 0.01 \pm 0.01$
    & $0.29 \pm 0.04~ ^{+0.03}_{-0.12}$ \\[1mm]
    Migration
    & $0.014 \pm 0.001 \pm 0.002$
    & $0.023 \pm 0.001 \pm 0.004$
    & $0.025 \pm 0.001 \pm 0.004$
    & $0.024 \pm 0.001 \pm 0.003$ \\[1mm]
    \hline
        & & & & \\[-2mm]
    Total non-$ZZ$
    & $0.17 \pm 0.01 \pm 0.03 $
    & $0.14 \pm 0.01 \pm 0.02 $
    & $0.08 \pm 0.01 \pm 0.01 $
    & $0.32 \pm 0.04~ ^{+0.03}_{-0.12} $ \\
    background & & & &\\[1mm]
    \hline 
    & & & & \\[-2mm]
    Expected
    & $0.48 \pm 0.01 \pm 0.07 $
    & $1.14 \pm 0.01 \pm 0.17 $
    & $1.03 \pm 0.01 \pm 0.15 $
    & $1.47 \pm 0.01 \pm 0.19 $ \\
    $t$-channel $Z/\gamma^{*}\mbox{ }Z/\gamma^{*}$ & & & &\\[1mm]
    \hline
    & & & & \\[-2mm]
    Expected $gg \to H$ 
    & $<0.001$
    & 0.001
    & 0.004
    & 0.002 \\
    $M_H = $ 125 GeV&& & &\\[1mm]
    Expected $ZH$
    & 0.003
    & 0.006
    & 0.010
    & 0.008 \\
    $M_H =$ 125 GeV&& & &\\[1mm]
    \hline
    & & & & \\[-2mm]
    Total Higgs boson 
    & 0.003
    & 0.007
    & 0.014 
    & 0.010 \\
    $M_H =$ 125 GeV & & & &\\[1mm]
    \hline
    & & & & \\[-2mm]
    Observed  & 0 & 1 & 2 & 2\\
    Events& & & &\\
    \hline \hline
  \end{tabular}
\end{table*}

\begin{table*}[!htb]
  \caption{\label{tab:emubkgd}Contributions from non-negligible 
    backgrounds in the $ee\mu\mu$ subchannels, plus expected signal and 
    number of observed events. Uncertainties are statistical followed by
    systematic.}
  \begin{tabular}{l|c|c|c} 

    \hline \hline
               & 0 CC & 1 CC & 2 CC \\ 
    \hline
               &      &      &      \\[-2mm]      
    Instrumental backg.
    & $0.11 \pm 0.01 \pm 0.03$
    & $0.21 \pm 0.01 \pm 0.04 $ 
    & $0.27 \pm 0.01 \pm 0.04 $ \\[1mm]
    $t\bar{t}$ 
    & $(0.2~^{+0.3}_{-0.1}\pm 0.6)\e{-2}$
    & $(1.0~^{+0.5}_{-0.3}\pm 0.2)\e{-2}$ 
    & $(0.3~^{+0.2}_{-0.1}\pm 0.3)\e{-2}$ \\[1mm] 
    Migration 
    & $(2.1~^{+0.9}_{-0.7}~^{+0.3}_{-1.0})\e{-3}$ 
    & $(5.0 \pm 0.8 ~^{+0.6}_{-1.4})\e{-3}$ 
    & $(4.8~^{+0.6}_{-0.5}~\pm 1.0)\e{-3}$ \\[1mm]     
    Cosmic rays 
    & $<0.001$ 
    & $<0.003$ 
    & $<0.006$ \\[1mm]
    \hline
               &      &      &      \\[-2mm]          
    Total non-$ZZ$
    & $0.12 \pm 0.01 \pm 0.03  $
    & $0.23 \pm 0.01 \pm 0.04 $
    & $0.27 \pm 0.01 \pm 0.04 $ \\
    background & & & \\[1mm]

    \hline
               &      &      &      \\[-2mm]      
    Expected 
    & $0.43 \pm 0.01 \pm 0.06 $
    & $2.37 \pm 0.02 \pm 0.28$
    & $4.13 \pm 0.03 \pm 0.49$ \\
    $t$-channel $Z/\gamma^{*}\mbox{ }Z/\gamma^{*}$ &  &      & \\[1mm]
    \hline
    & & & \\[-2mm]
    Expected $gg \to H$   & $<0.001$ &  0.002 & 0.007 \\
    $M_H =$ 125 GeV& & & \\[1mm]
    & & & \\[-2mm]
    Expected $ZH$ & 0.001 & 0.015 & 0.036 \\
    $M_H =$ 125 GeV& & &\\[1mm]
    \hline
    & & & \\[-2mm]
    Total Higgs boson & 0.002 & 0.017 & 0.043 \\
    $M_H =$ 125 GeV&& &\\[1mm]
    \hline
    Observed Events & 2 & 1 & 2 \\
    \hline \hline
  \end{tabular}
\end{table*}

\section{\label{sec:syst}Systematic Uncertainties}

The following factors contribute to the systematic uncertainty on this measurement.  
We assess a 1\%  trigger efficiency uncertainty.  Lepton identification uncertainties are calculated 
by studying $Z \to \ell \ell$ events;  lepton identification uncertainties of 3.7\% per CC and 
EC electron, 6\% per ICR electron, and 3.2\% per muon are used.
There is a 10\%--50\% systematic uncertainty on the instrumental background expectation in the various 
final states that is due to observed variations in $P_{j \ell}$ when changing selection
requirements for the di-jet sample as well as limited statistics in the data samples used.
We assign 20\% uncertainty to the $t\bar{t}$ background. This covers
uncertainty on the theoretical production rate of 7\% for $m_{\rm top}=172$~GeV~\cite{Moch}, plus
variation in the cross section due to uncertainty on the top quark mass, and also that on the rate 
at which the $b$ quark from top quark decays is misidentified as an isolated lepton.
We estimate a PDF uncertainty of 2.5\% on all MC samples. We assign a 7.1\% uncertainty
on the $ZZ$ cross section used to estimate the 
migration background and the $ZZ$ background to the Higgs boson search.
A systematic uncertainty of 6.1\% is assessed on the luminosity measurement \cite{newlumi}.
We assess a systematic uncertainty on the $ZZ$ $p_T$  distribution by reweighting the {\sc pythia} 
$ZZ$ $p_T$ to match a distribution derived from {\sc sherpa} MC \cite{sherpa}.  The $ZZ$ $p_T$ 
systematic is between 1\% and 7\% for signal $t$-channel $ZZ$ events, but has up 
to a 40\% effect on the migration background.  
We also assess systematic uncertainties on the muon and electron 
energy resolution \cite{mumom}, which lead to an uncertainty on the cross section measurements and Higgs boson 
production limits of less than 2\%.
For the Higgs boson search, we assess a theoretical uncertainty on the expected gluon fusion 
and $ZH$ associated cross sections of 10.9\% and 6.2\%, respectively \cite{higgsxsec1,higgsxec2}.

\begin{table}[!htb]
\begin{center}
  \caption{\label{tab:mubkgd}Contributions from non-negligible 
    backgrounds in the $\mu\mu\mu\mu$ channel, plus expected $t$-channel $ZZ$ and Higgs boson signal and 
    number of observed events. Uncertainties are statistical followed by
    systematic.}
  \begin{tabular}{l|c} 
    \hline \hline
               & Number of Events\\
    \hline
        & \\[-2mm]                
    Instrumental backg.
    & $0.12 \pm 0.01~ ^{+0.07}_{-0.05}$ \\[1mm]
      & \\[-2mm]                
    Migration 
    & $(0.34 \pm 0.02~ ^{+0.07}_{-0.04})\e{-1}$ \\[1mm]
         & \\[-2mm]      
    Cosmic rays
    &  $<$0.01 \\[1mm]
    \hline 
        & \\[-2mm]      
    Total non-$ZZ$
    & $0.15 \pm 0.01~ ^{+0.07}_{-0.05}$ \\
    background & \\[1mm]
    \hline 
        & \\[-2mm]      
    Expected
    & $4.26 \pm 0.02  \pm 0.43$ \\
    $t$-channel $Z/\gamma^{*}\mbox{ }Z/\gamma^{*}$ &\\[1mm]
    \hline
    &  \\[-2mm]
    Expected $gg \to H$  & 0.007 \\
    $M_H =$ 125 GeV&\\[1mm]
    &  \\[-2mm]
    Expected $ZH$ & 0.033 \\
    $M_H =$ 125 GeV&\\[1mm]
    \hline
    &  \\[-2mm]
    Total Higgs boson & 0.040 \\
    $M_H =$ 125 GeV&\\[1mm]
    \hline
    Observed Events & 3 \\
    \hline \hline

  \end{tabular}
  \end{center}
\end{table}

\newpage
\section{\label{sec:results} Cross Section measurement}

The data are used to measure the production cross section 
$p\bar{p} \to ZZ$ at $\sqrt{s} = 1.96$~TeV.  
The integrated luminosities analyzed for the three channels are 9.8, 9.6, and 9.6 fb$^{-1}$ for the $eeee$, $ee\mu\mu$, 
and $\mu \mu \mu \mu$ channels, respectively.  
A summary of the signal and background event
expectations are included in Tables~\ref{tab:embkgd}--\ref{tab:mubkgd} for the three channels.

We observe five $eeee$ candidate events, five $ee \mu\mu$ candidate events, and 
three $\mu\mu\mu\mu$ candidate events, for 13 data events total, with a total
of $16.8 \pm 1.9 \thinspace \mbox{(stat+syst+lumi)}$ 
expected events.

\begin{table*}[!htb]
  \caption{\label{tab:acceptance_4e}  Acceptance $\times$ efficiency for the $eeee$ subchannels,
    for $ZZ \to ee ee$ and $ZZ \to ee \tau \tau$ decays.
  Uncertainties are statistical followed by systematic.}
  \begin{tabular}{l|c|c} 
    \hline \hline
    Channel        & $eeee$ & $ee\tau\tau$ \\ 
\hline
       2 CC, 2 EC & $0.025 \pm 0.001 \pm 0.004$ & $0.0002 \pm 0.0001 \pm 0.0001$ \\
       3 CC, 1 EC & $0.059 \pm 0.001 \pm 0.011$ & $0.0006 \pm 0.0001 \pm 0.0001$ \\
       4 CC       & $0.053 \pm 0.001 \pm 0.009$ & $0.0007 \pm 0.0001 \pm 0.0001$ \\
$\geq$ 2 CC, 1 ICR & $0.076 \pm 0.001 \pm 0.012$ & $0.0007 \pm 0.0001 \pm 0.0001$ \\
   \hline \hline
  \end{tabular}
\end{table*}

\begin{table*}[!htb]
  \caption{\label{tab:acceptance_2e2mu}  Acceptance $\times$ efficiency for the $ee\mu\mu$ subchannels, 
  for $ZZ \to ee \mu \mu$, $ZZ \to ee \tau \tau$, and $ZZ \to \mu \mu \tau \tau$ decays.  
  Uncertainties are statistical followed by systematic.}
  \begin{tabular}{l|c|c|c} 
    \hline \hline
    Channel        & $ee\mu\mu$ & $ee\tau\tau$ & $\mu \mu \tau\tau$ \\ 
\hline
       0 CC & $0.011 \pm 0.001 \pm 0.001$ & $0.0001 \pm 0.0001 \pm 0.0001$ & $0.0002 \pm 0.0001 \pm 0.0001$ \\
       1 CC & $0.063 \pm 0.001 \pm 0.007$ & $0.0007 \pm 0.0001 \pm 0.0001$ & $0.0007 \pm 0.0001 \pm 0.0001$  \\ 
       2 CC & $0.110 \pm 0.001 \pm 0.012$ & $0.0014 \pm 0.0001 \pm 0.0002$ & $0.0019 \pm 0.0001 \pm 0.0002$  \\
   \hline \hline
  \end{tabular}
\end{table*}

\begin{table*}[!htb]
  \caption{\label{tab:acceptance_4mu}  Acceptance $\times$ efficiency for the $\mu \mu \mu \mu$ channel, for $ZZ \to \mu \mu \mu \mu$
  and $ZZ \to \mu \mu \tau \tau$ decays. Uncertainties are statistical followed by systematic.}
  \begin{tabular}{c|c} 
    \hline \hline
    $\mu \mu \mu\mu$ & $\mu \mu \tau\tau$ \\ 
\hline
       $0.224 \pm 0.002 \pm 0.022$ & $0.0032 \pm 0.0002 \pm 0.0003$ \\
   \hline \hline
  \end{tabular}
\end{table*}

A negative log-likelihood function is constructed by taking as input the expected signal acceptance,
the number of expected background events, and the number of observed events in each of the subchannels.
The signal acceptance times efficiency for each channel are shown in Tables~\ref{tab:acceptance_4e}--\ref{tab:acceptance_4mu}.  
The branching ratio for each channel is calculated using the relevant $Z$ boson branching ratios from Ref.~\cite{pdg}.
The cross section, $\sigma$, is varied to minimize the negative log-likelihood, which gives
$\sigma(p\bar{p} \to Z/\gamma^*Z/\gamma^*)= 1.26 ^{+0.44}_{-0.36} \thinspace \mbox{(stat)}  ^{+0.17}_{-0.15} \thinspace \mbox{(syst)} \pm 0.08 \thinspace \mbox{(lumi)}$~pb for 
$M(Z/\gamma^{*}) > 30$~GeV.  We then 
calculate the ratio of $\sigma(p\bar{p} \to Z/\gamma^*Z/\gamma^*)$ to 
$\sigma(p\bar{p} \to ZZ)$ for this mass region using {\sc mcfm} \cite{MCFM}, and from this correction determine 
the $p\bar{p} \to ZZ$ cross section to be  
$1.05 ^{+0.37}_{-0.30} \thinspace \mbox{(stat)}  ^{+0.14}_{-0.12} \thinspace \mbox{(syst)} \pm 0.06\thinspace \mbox{(lumi)}$~pb.  
We combine this measurement with the $p\bar{p} \to ZZ$ cross section measured in the $\ell^{+}\ell^{-}\nu\bar{\nu}$ final state using data 
from the D0 detector \cite{d0_zz_2l2nu}, giving a total combined $p\bar{p} \to ZZ$ cross section of 
$1.32 ^{+0.29}_{-0.25} \thinspace \mbox{(stat)}  \pm 0.12 \thinspace \mbox{(syst)} \pm 0.04\thinspace \mbox{(lumi)}$~pb.
The measured cross section values are
consistent with the SM expectation of $1.43 \pm 0.10$ pb \cite{MCFM}.

\section{Higgs Boson Production Limits \label{higgsSection}}
The main Higgs boson production mechanisms that can result in four final state charged leptons are 
gluon fusion and $ZH$ associated production.  

For Higgs boson events produced through gluon fusion, final states with four charged leptons arise from 
the decay $H \to ZZ$, where both $Z$ bosons then decay leptonically.  As all of the decay products 
of the Higgs boson in this decay are well measured, the best discriminating variable between 
the gluon fusion Higgs boson signal and the backgrounds is the four-lepton invariant mass.

In the case of associated $ZH$ production, two of the leptons in each event can come from the decay of 
the associated $Z$ boson, so Higgs decay modes with two or more final state leptons will contribute to our signal. 
The majority of the $ZH$ signal arises from $H \to \tau^{+}\tau^{-}$, $H \to WW$, and $H \to ZZ$ decays.  We expect large 
\met~in these events, due to the neutrinos from the $\tau$ and $W$ boson decays, as well as in events where one 
$Z$ boson from the $H \to ZZ$ decays to neutrinos.

\begin{figure}[!htb]
\centering
\includegraphics[width=\columnwidth]{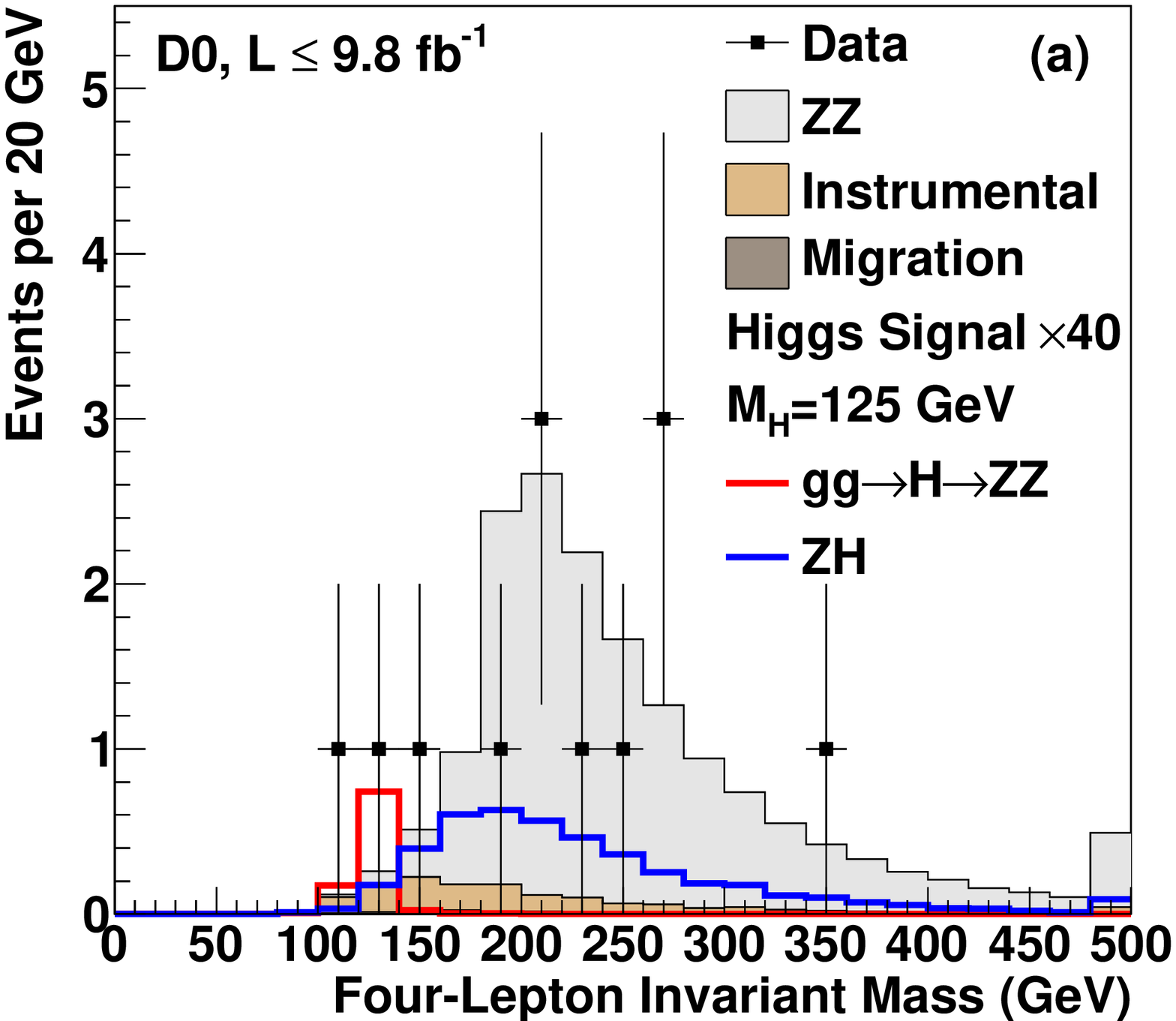}
\includegraphics[width=\columnwidth]{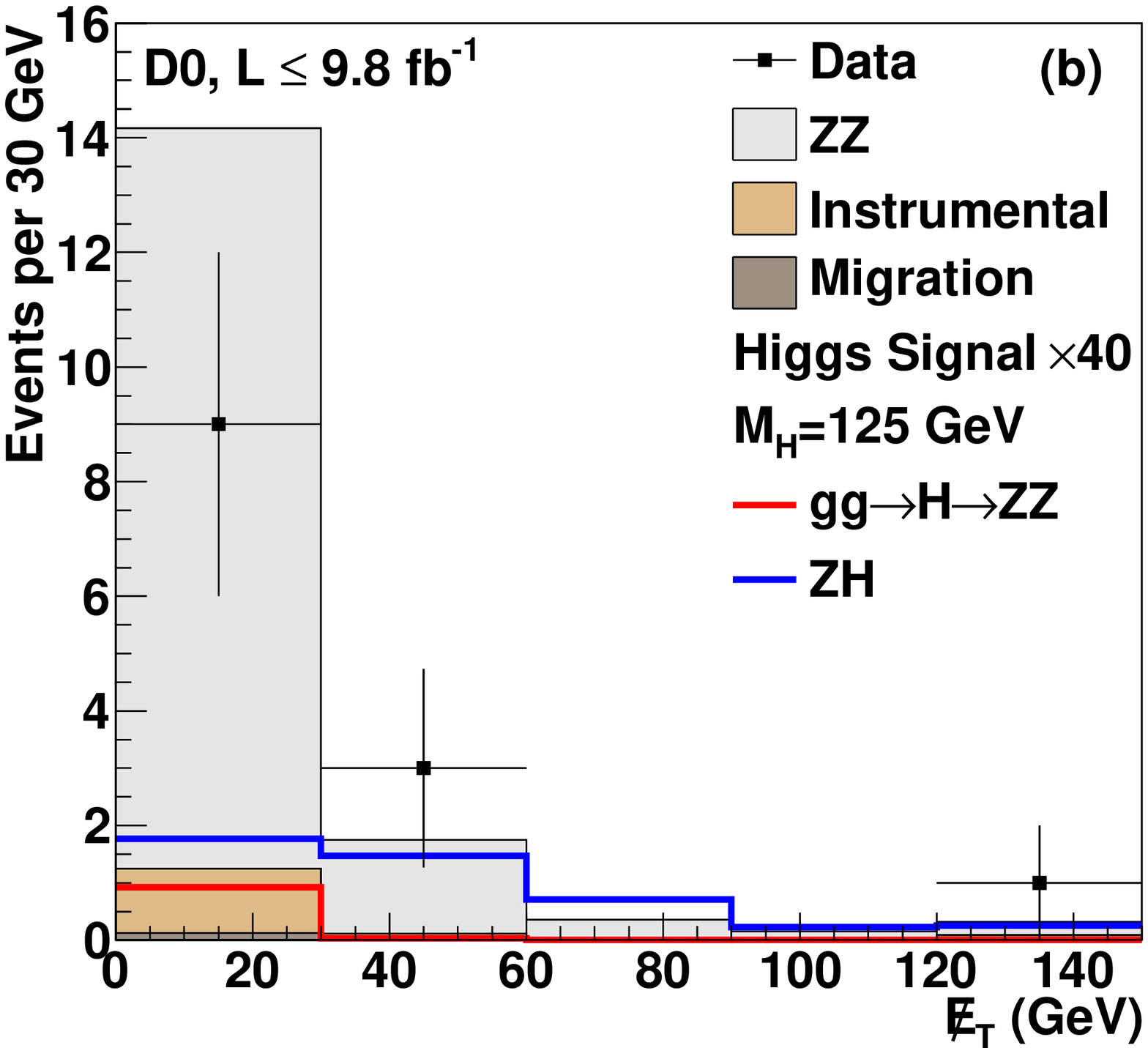}
\caption{ \label{fig:fourmass} Distributions of (a) the four lepton invariant mass and (b) the \met~ in data, and of expected signal and background.
The Higgs boson signal for $M_H$ of 125 GeV is shown scaled by a factor of 40.}
\end{figure}

We therefore set limits on SM Higgs boson production using the four-lepton invariant mass and the \met.  
The four-lepton mass and \met~are shown in Fig.~\ref{fig:fourmass}, with the expected Higgs boson 
signal distributions for a Higgs boson mass, $M_H$, of 125 GeV.  Additional differential distributions 
are provided in Appendix~\ref{sec:kinematics}.  The expected yields for each production and decay 
mode for each Higgs boson mass considered are shown in Table~\ref{tab:decaytable}.  
For events with $\mbox{\met}< 30$ GeV, the four-lepton mass is used to discriminate the Higgs boson signal 
from all backgrounds;  in events with $\mbox{\met} \geq  30$ GeV, the \met~is used. For the Higgs boson 
search, the $t$-channel $Z/\gamma^*\mbox{ }Z/\gamma^*$ background is fixed to the SM expectation.

\begin{table*}[!htb]
  \caption{\label{tab:decaytable}  Expected numbers of Higgs boson events for each mass point for the given production and decay 
mode.  The $H \to \gamma \gamma$, $H \to \mu \mu$, and $H \to Z\gamma$ contributions are summed together in the $H \to \mbox{other}$ decays 
column.}
 \begin{tabular}{c|c|cccc|c} 
    \hline \hline
  $M_H$ & $gg \to H$    & \multicolumn{4}{c|}{$q\bar{q} \to ZH$}                                         & Total \\
  (GeV) & $H \to ZZ$    & $H \to WW$    & $H \to ZZ$    & $H \to \tau \tau$            & $H \to \mbox{other}$ & \\
\hline
115	&	0.009	&	0.016	&	0.013	&	0.060	&	0.008	&	0.106	\\
120	&	0.013	&	0.026	&	0.017	&	0.052	&	0.006	&	0.113	\\
125	&	0.024	&	0.040	&	0.024	&	0.043	&	0.005	&	0.137	\\
130	&	0.049	&	0.058	&	0.039	&	0.035	&	0.004	&	0.184	\\
135	&	0.090	&	0.066	&	0.047	&	0.025	&	0.003	&	0.232	\\
140	&	0.138	&	0.077	&	0.055	&	0.018	&	0.003	&	0.291	\\
145	&	0.185	&	0.088	&	0.061	&	0.013	&	0.002	&	0.348	\\
150	&	0.210	&	0.092	&	0.059	&	0.008	&	0.001	&	0.371	\\
155	&	0.196	&	0.099	&	0.049	&	0.004	&	0.001	&	0.348	\\
160	&	0.112	&	0.100	&	0.026	&	0.002	&	0.000	&	0.240	\\
165	&	0.059	&	0.097	&	0.012	&	0.001	&	0.000	&	0.169	\\
170	&	0.062	&	0.088	&	0.012	&	0.000	&	0.000	&	0.162	\\
175	&	0.082	&	0.086	&	0.015	&	0.000	&	0.000	&	0.183	\\
180	&	0.148	&	0.078	&	0.027	&	0.000	&	0.000	&	0.254	\\
185	&	0.348	&	0.068	&	0.063	&	0.000	&	0.000	&	0.478	\\
190	&	0.440	&	0.058	&	0.077	&	0.000	&	0.000	&	0.575	\\
195	&	0.467	&	0.051	&	0.082	&	0.000	&	0.000	&	0.600	\\
200	&	0.468	&	0.046	&	0.083	&	0.000	&	0.000	&	0.597	\\
   \hline \hline
  \end{tabular}
\end{table*}

We find no evidence of SM Higgs boson production and proceed to 
set limits.  We consider potential $M_H$ values between 115 and 
200 GeV, in 5 GeV increments.
We calculate limits on the SM Higgs boson production cross section
using a modified frequentist approach~\cite{bib-wade1,bib-wade2,bib-wade3}.
A log-likelihood ratio (LLR) test statistic
is formed using the Poisson probabilities for estimated
background yields, the expected signal acceptance, and the 
number of observed events for each considered Higgs boson mass hypothesis.
The confidence levels are derived by integrating the
LLR distribution in pseudo-experiments using both the signal-plus-background hypothesis
({\it CL}$_{s+b}$) and the background-only
hypothesis ({\it CL}$_b$). The excluded production cross section is
taken to be the cross section for which the confidence
level for signal, {\it CL}$_s={\mbox{\it CL}}_{s+b}/{\mbox{\it CL}}_b$, is less than or equal to $0.05$.

\begin{figure}[b]
    \includegraphics[width=0.45\textwidth]{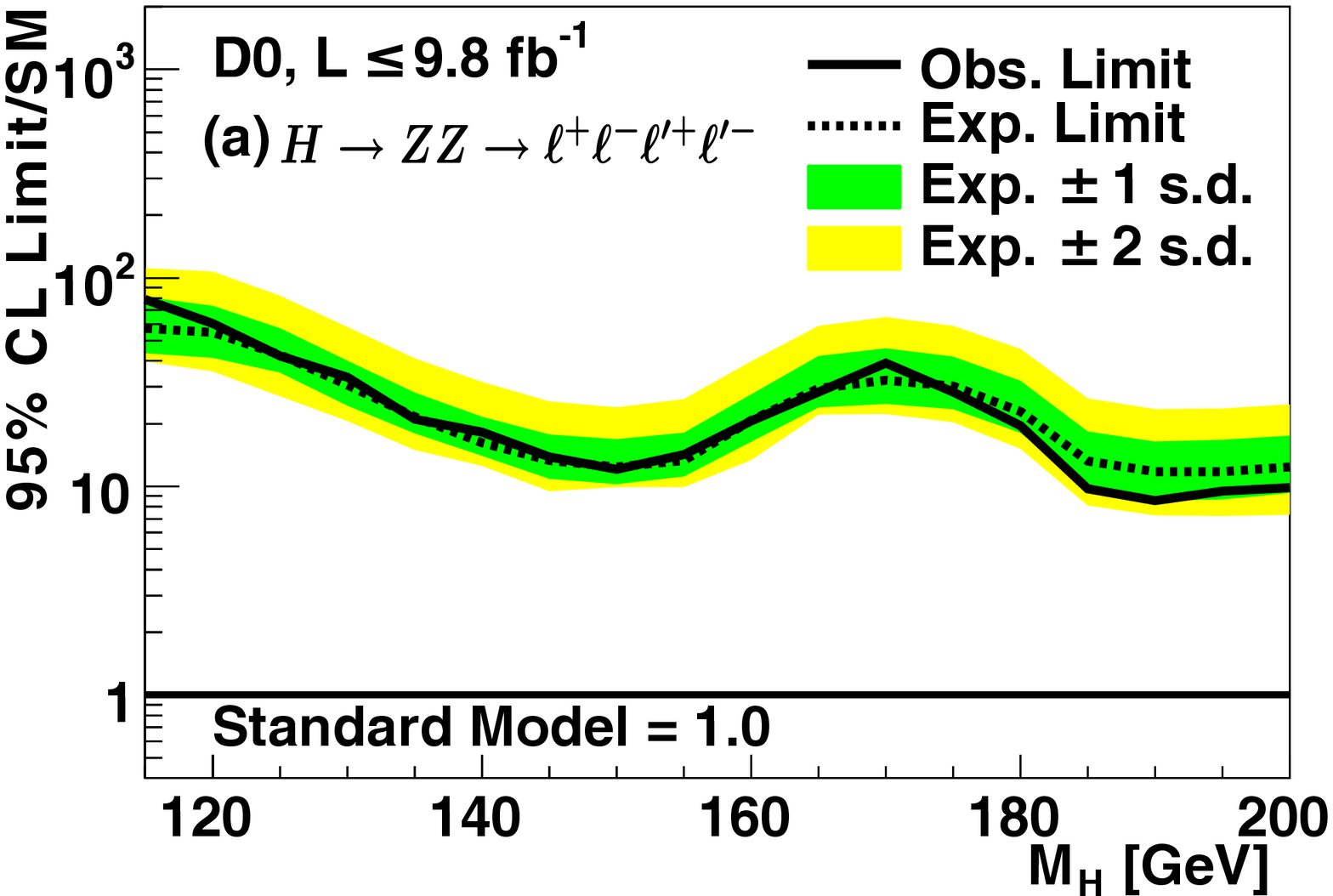}
    \includegraphics[width=0.45\textwidth]{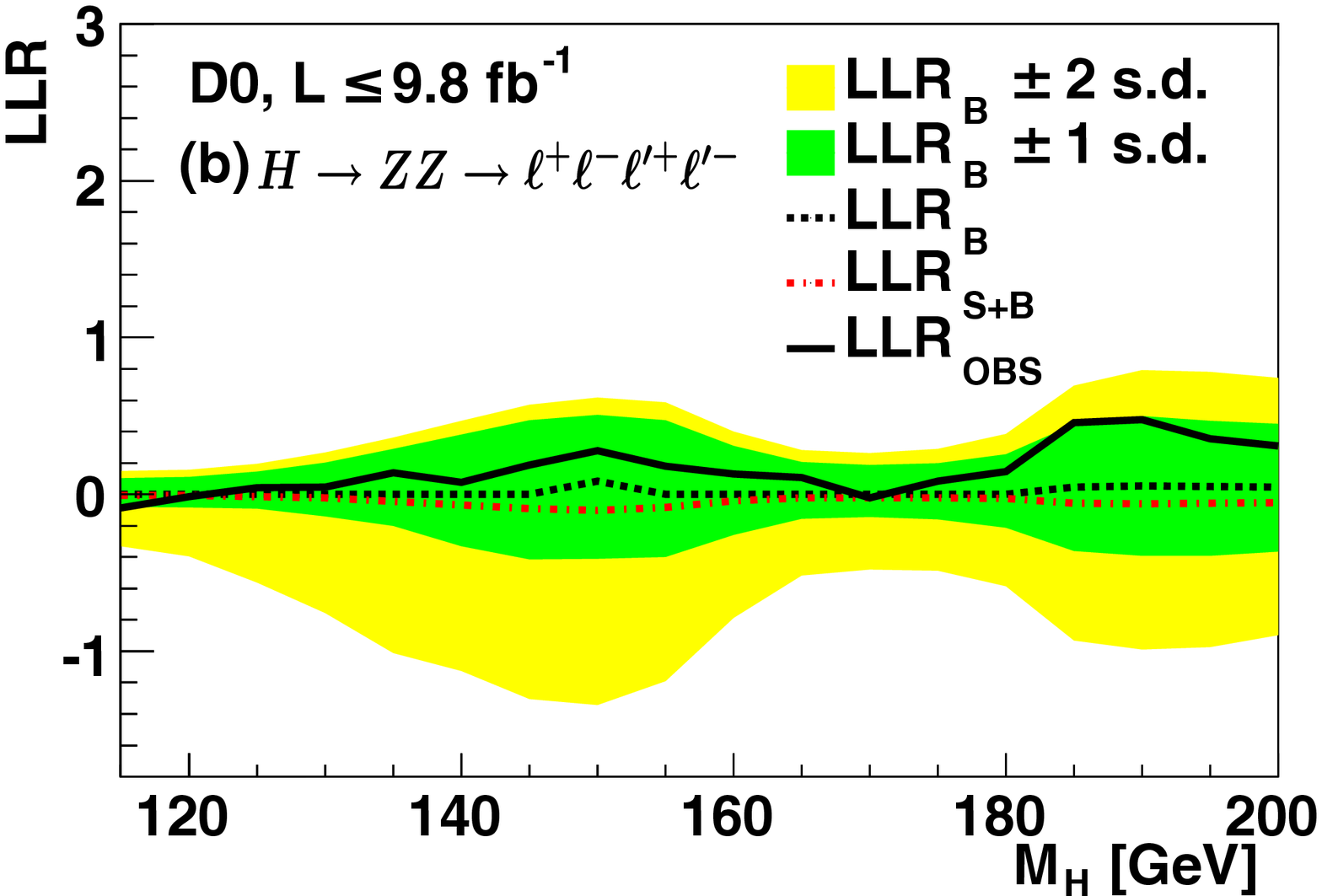}
  \caption{ The (a) expected and observed 95\% C.L. upper limits on the SM Higgs boson production cross section relative to 
the value expected in the SM, and the (b) log-likelihood ratio for all four lepton channels combined.}
   \label{fig:limit}
\end{figure}

The calculated limits are listed in Table~\ref{tab:limits}.  At $M_H=$ 125 GeV, we expect 
to set a limit of 42.8 times the SM cross section at the 95\% C.L., and observe a limit of 42.3 times the SM cross section.  
The limits vs. $M_H$ are shown in Fig.~\ref{fig:limit}, along with the associated LLR 
distribution.

\begin{table}[b]
  \caption{\label{tab:limits}  Expected and observed 95\% C.L. upper limits on the SM Higgs boson production cross section relative to 
the value expected in the SM.}
  \begin{tabular}{c|c|c} 
\hline
\hline
$M_{H}$ (GeV)	&	Expected	&	Observed	\\
\hline
115	&	57.3	&	78.9	\\
120	&	54.9	&	60.6	\\
125	&	42.8	&	42.3	\\
130	&	30.6	&	33.5	\\
135	&	21.5	&	21.0	\\
140	&	16.2	&	18.2	\\
145	&	13.4	&	13.9	\\
150	&	12.4	&	12.1	\\
155	&	13.4	&	14.2	\\
160	&	20.8	&	20.6	\\
165	&	29.6	&	28.3	\\
170	&	32.3	&	39.0	\\
175	&	30.4	&	28.4	\\
180	&	22.9	&	19.6	\\
185	&	13.3	&	9.7	\\
190	&	11.8	&	8.6	\\
195	&	11.8	&	9.5	\\
200	&	12.4	&	9.9	\\
\hline
\hline
  \end{tabular}
\end{table}

\section{Conclusions}
We have measured the production cross section for $p\bar{p} \to Z/\gamma^{*} \mbox{ } Z/\gamma^{*}$ with $M(Z/\gamma^{*}) > 30$ GeV to be  
$1.26 ^{+0.44}_{-0.36} \thinspace \mbox{(stat)}  ^{+0.17}_{-0.15} \thinspace \mbox{(syst)} \pm 0.08 \thinspace \mbox{(lumi)}$ pb.  We correct this 
measurement by the 
expected ratio of $\sigma(p\bar{p} \to Z/\gamma^*Z/\gamma^*)$ to 
$\sigma(p\bar{p} \to ZZ)$ for this mass region and obtain a  $p\bar{p} \to ZZ$  cross section of
$1.05 ^{+0.37}_{-0.30} \thinspace \mbox{(stat)}  ^{+0.14}_{-0.12} \thinspace \mbox{(syst)} \pm 0.06\thinspace \mbox{(lumi)}$~pb. 
We also searched for the Higgs boson in the four lepton 
final state, assuming that the $t$-channel $ZZ$ pair is produced with the cross section predicted by the 
SM.  At $M_{H} = 125$ GeV, we expect a limit of 42.8 times the SM cross section, and set a limit of 42.3 times the SM cross section at the 95\% C.L. 

%
We thank the staffs at Fermilab and collaborating institutions,
and acknowledge support from the
DOE and NSF (USA);
CEA and CNRS/IN2P3 (France);
MON, NRC KI and RFBR (Russia);
CNPq, FAPERJ, FAPESP and FUNDUNESP (Brazil);
DAE and DST (India);
Colciencias (Colombia);
CONACyT (Mexico);
NRF (Korea);
FOM (The Netherlands);
STFC and the Royal Society (United Kingdom);
MSMT and GACR (Czech Republic);
BMBF and DFG (Germany);
SFI (Ireland);
The Swedish Research Council (Sweden);
and
CAS and CNSF (China).

\appendix

\section{Differential Distributions \label{sec:kinematics}}
\begin{figure*}[!htb]
\centering
\includegraphics[width=0.40\textwidth]{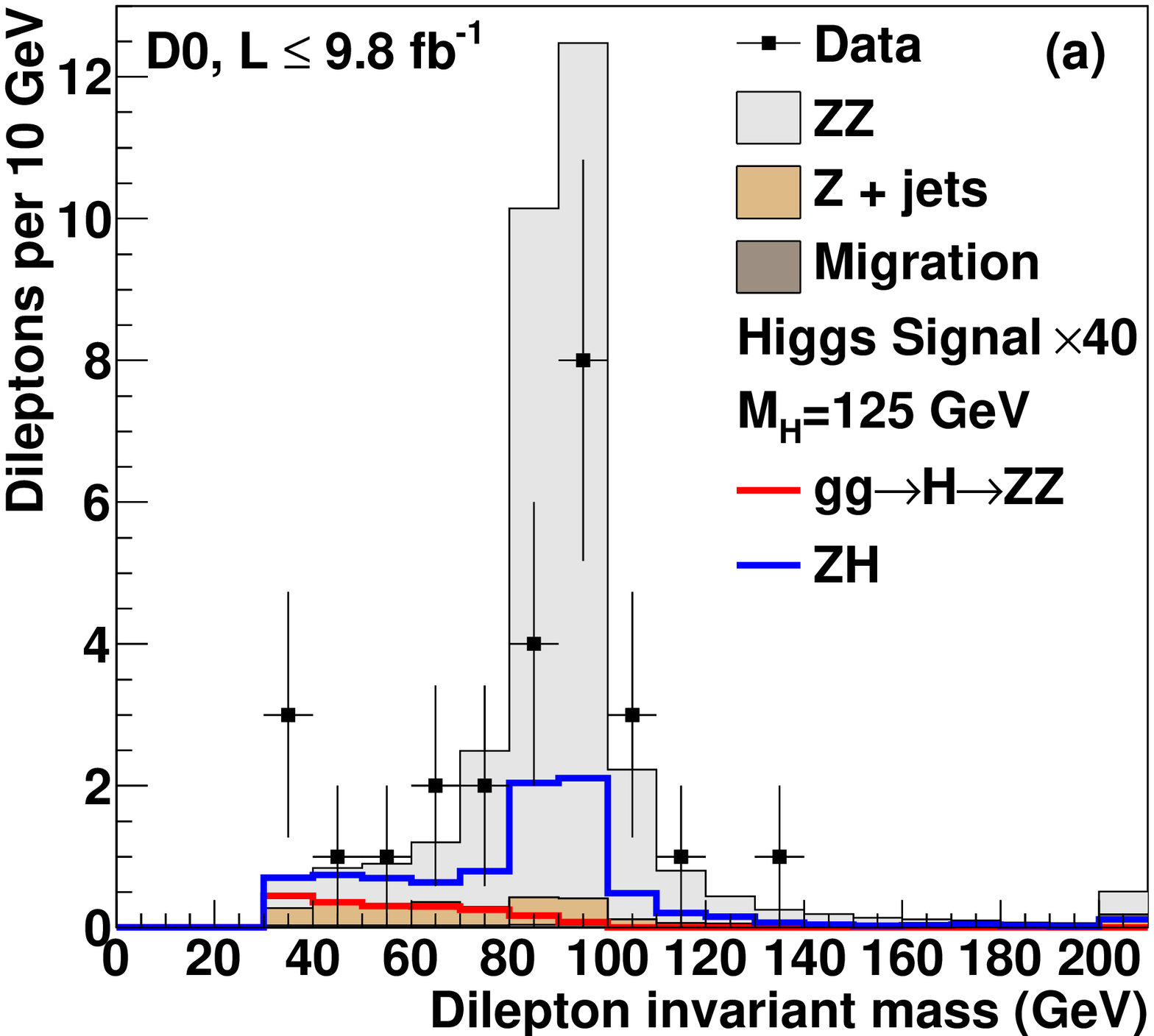}
\includegraphics[width=0.40\textwidth]{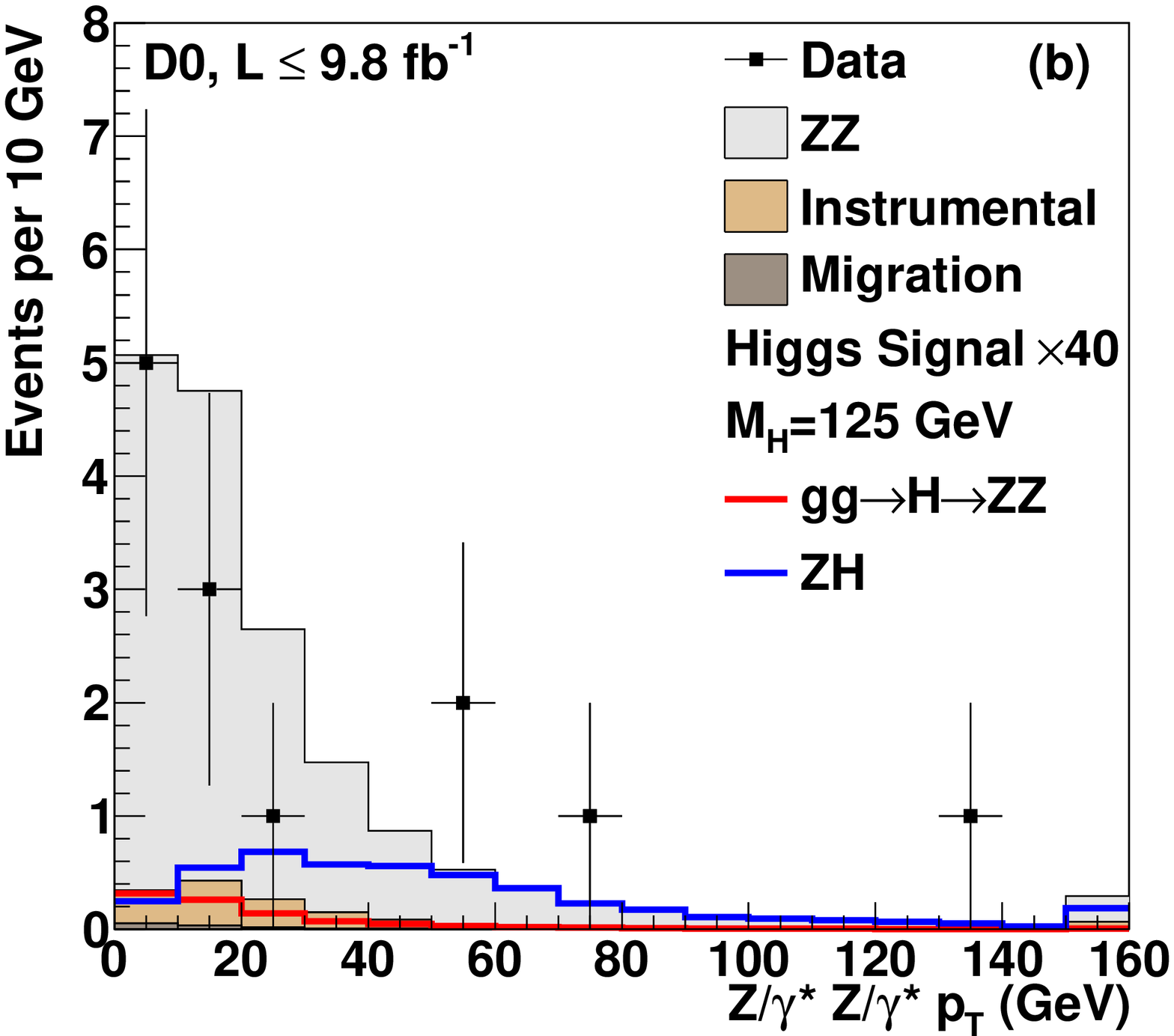}
\caption{ \label{fig:dileptonmasszzpt} Distributions of (a) the dilepton invariant mass and (b) the transverse momentum of the four-lepton system in 
data, expected signal and background.  There are two entries per event in the dilepton invariant mass distribution.
The Higgs boson signal for $M_H$ of 125 GeV is shown scaled by a factor of 40.}
\end{figure*}

\begin{figure*}[!htb]
\centering
\includegraphics[width=0.40\textwidth]{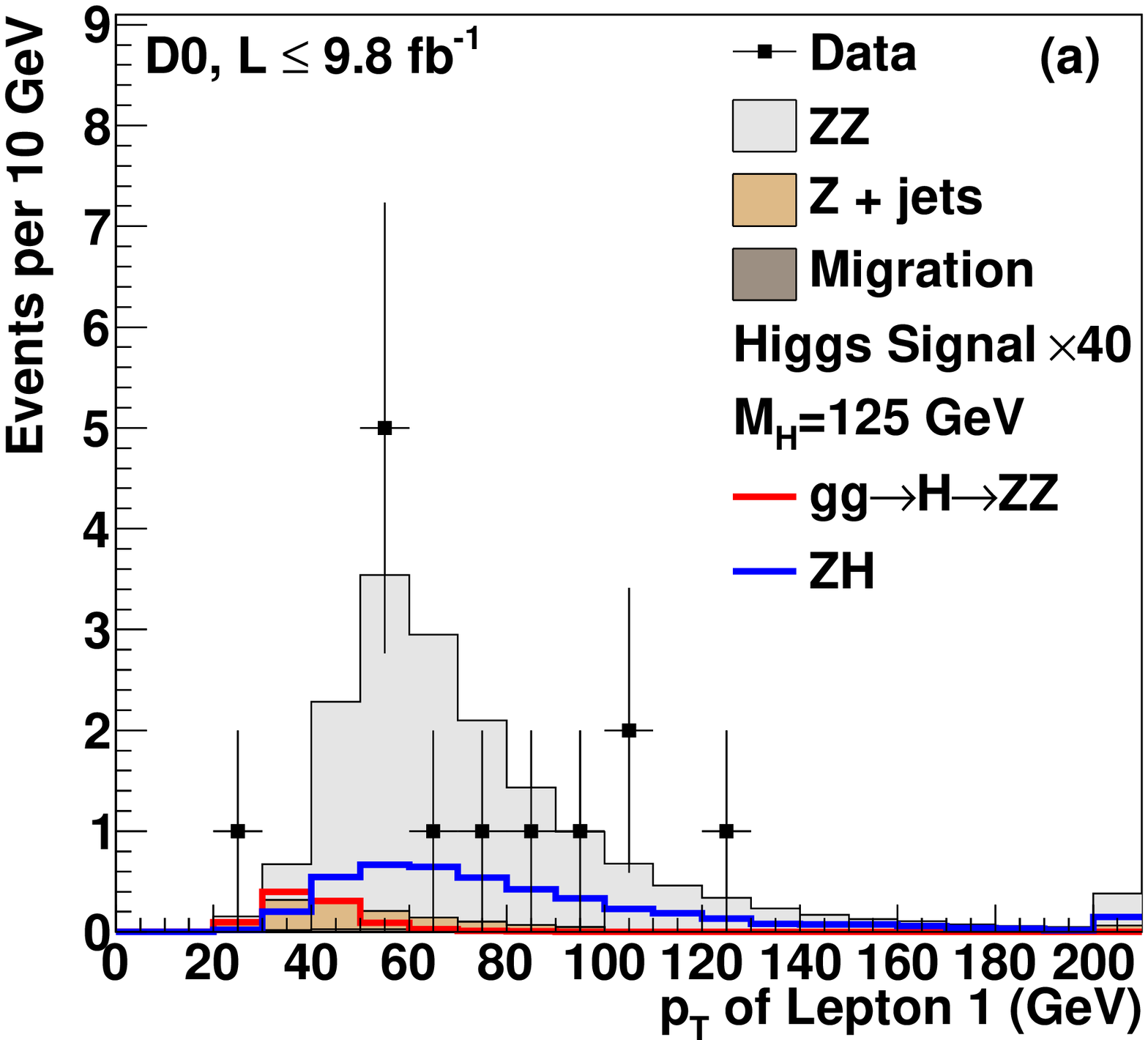}
\includegraphics[width=0.40\textwidth]{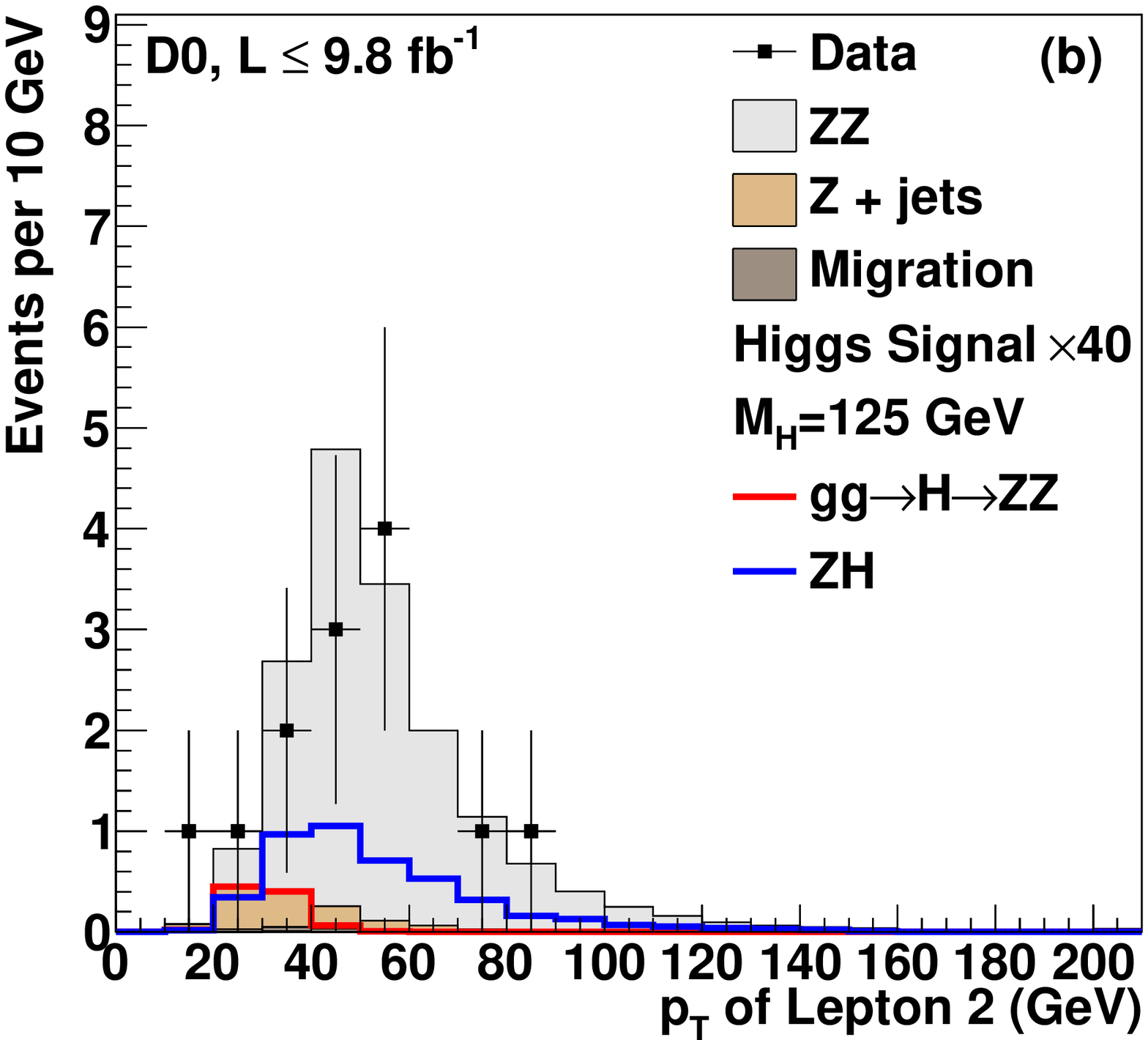}\\
\includegraphics[width=0.40\textwidth]{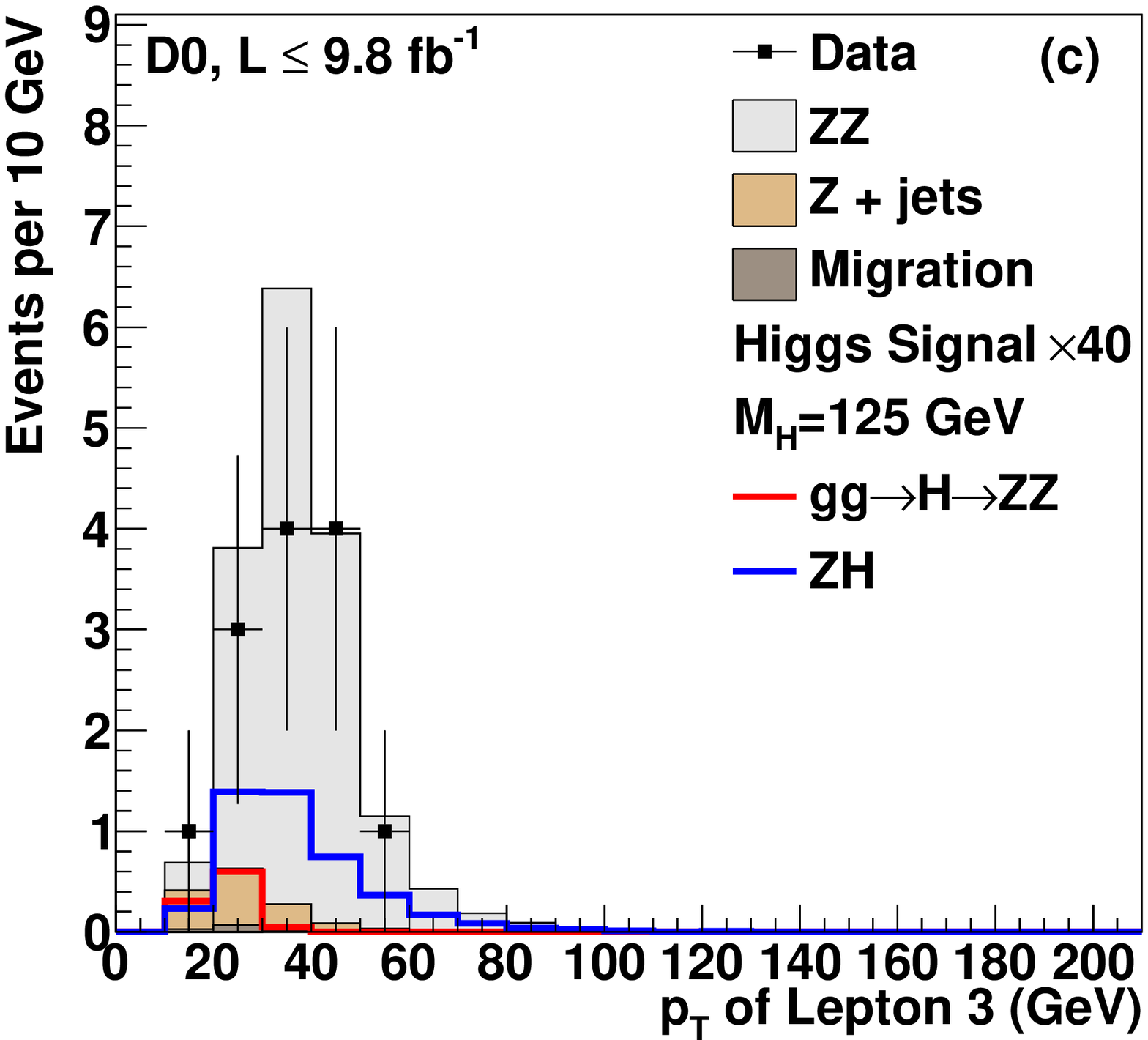}
\includegraphics[width=0.40\textwidth]{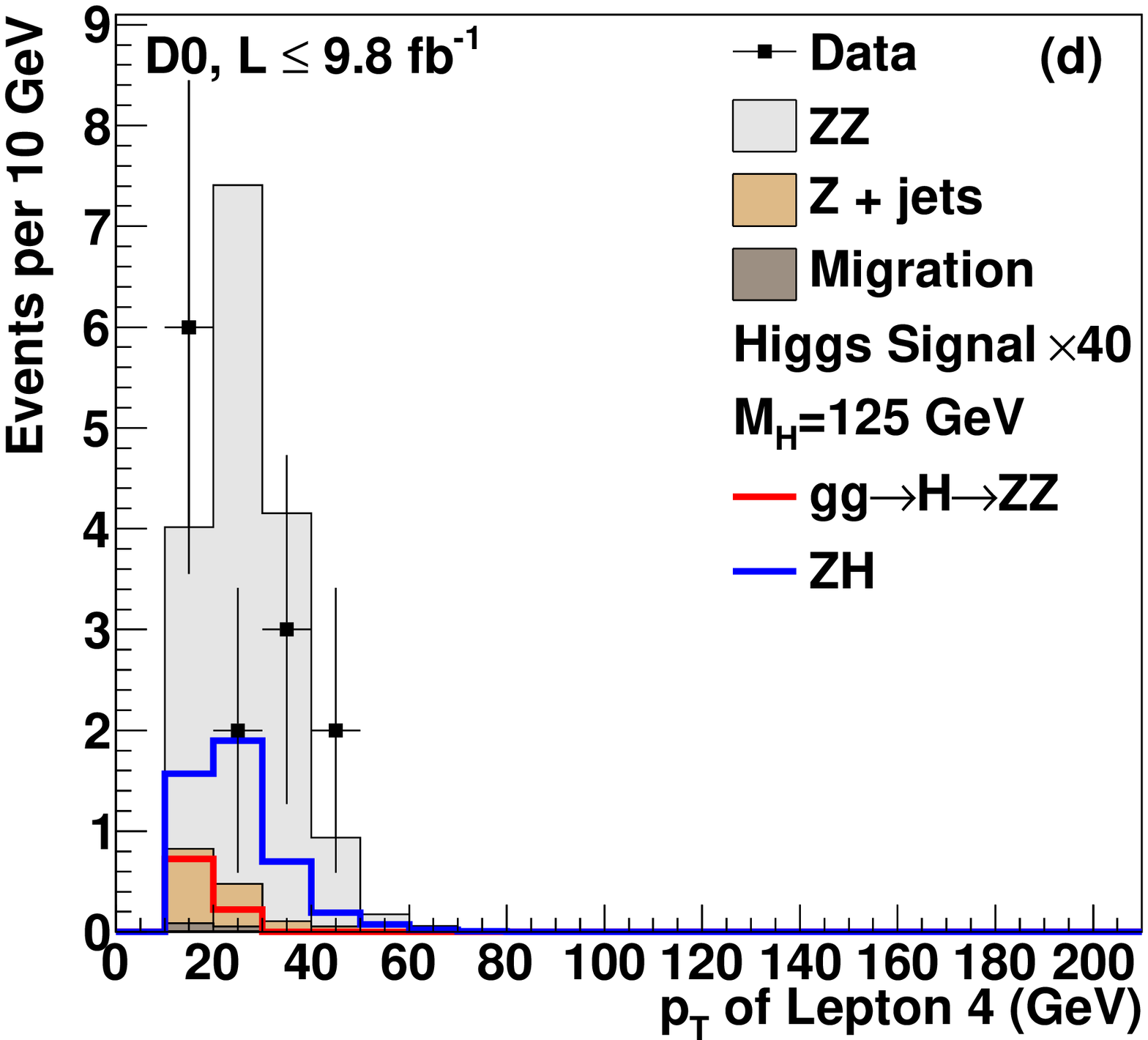}
\caption{ \label{fig:leptonpt} Distributions of the transverse momentum in data, expected signal, and backgrounds for the (a) highest-$p_T$, (b) second-highest-$p_T$, (c) third-highest-$p_T$, and (d) lowest-$p_T$ leptons in each event.  
The Higgs boson signal for $M_H$ of 125 GeV is shown scaled by a factor of 40.}
\end{figure*}

\begin{figure*}[!htb]
\centering
\includegraphics[width=0.40\textwidth]{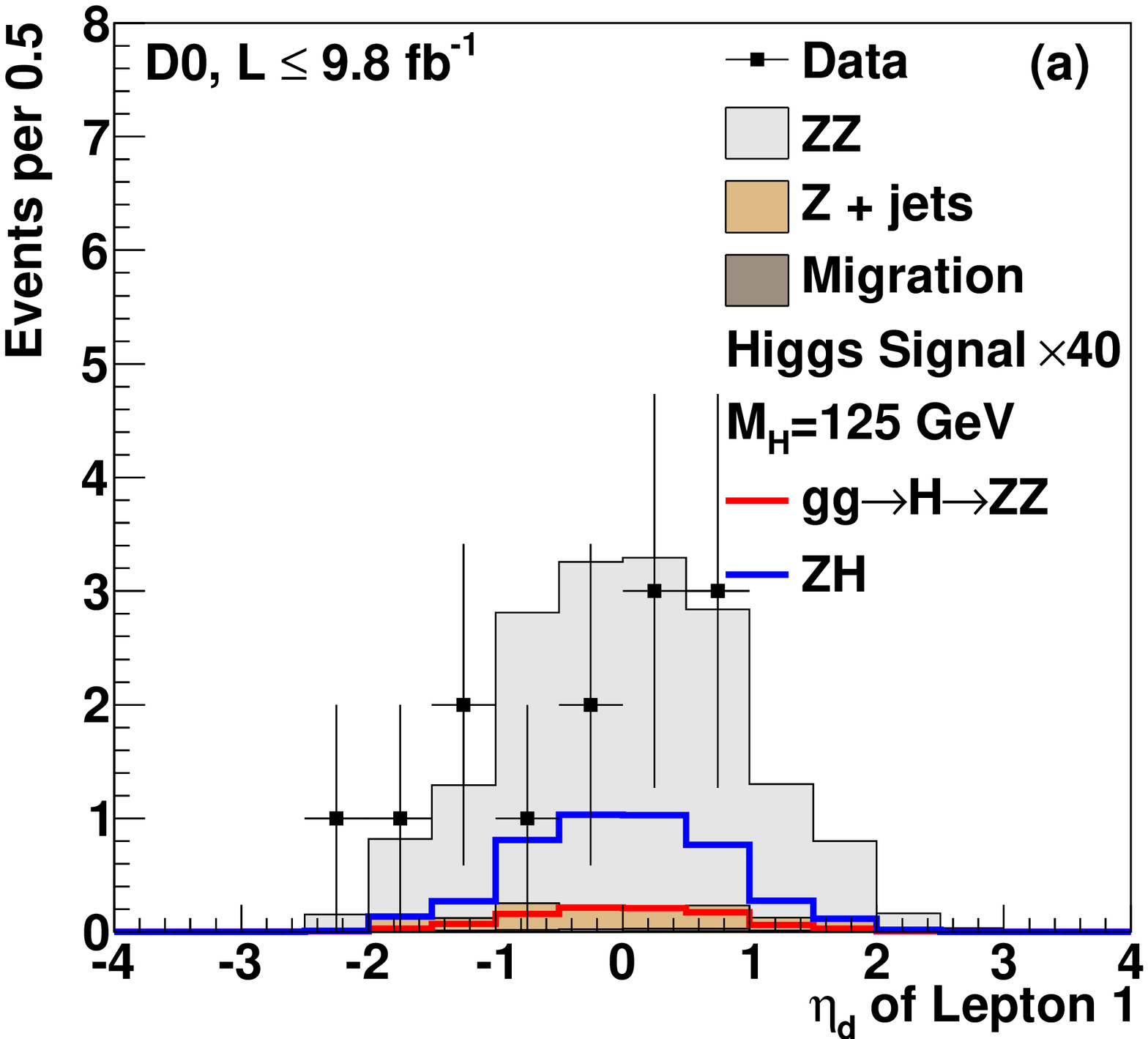}
\includegraphics[width=0.40\textwidth]{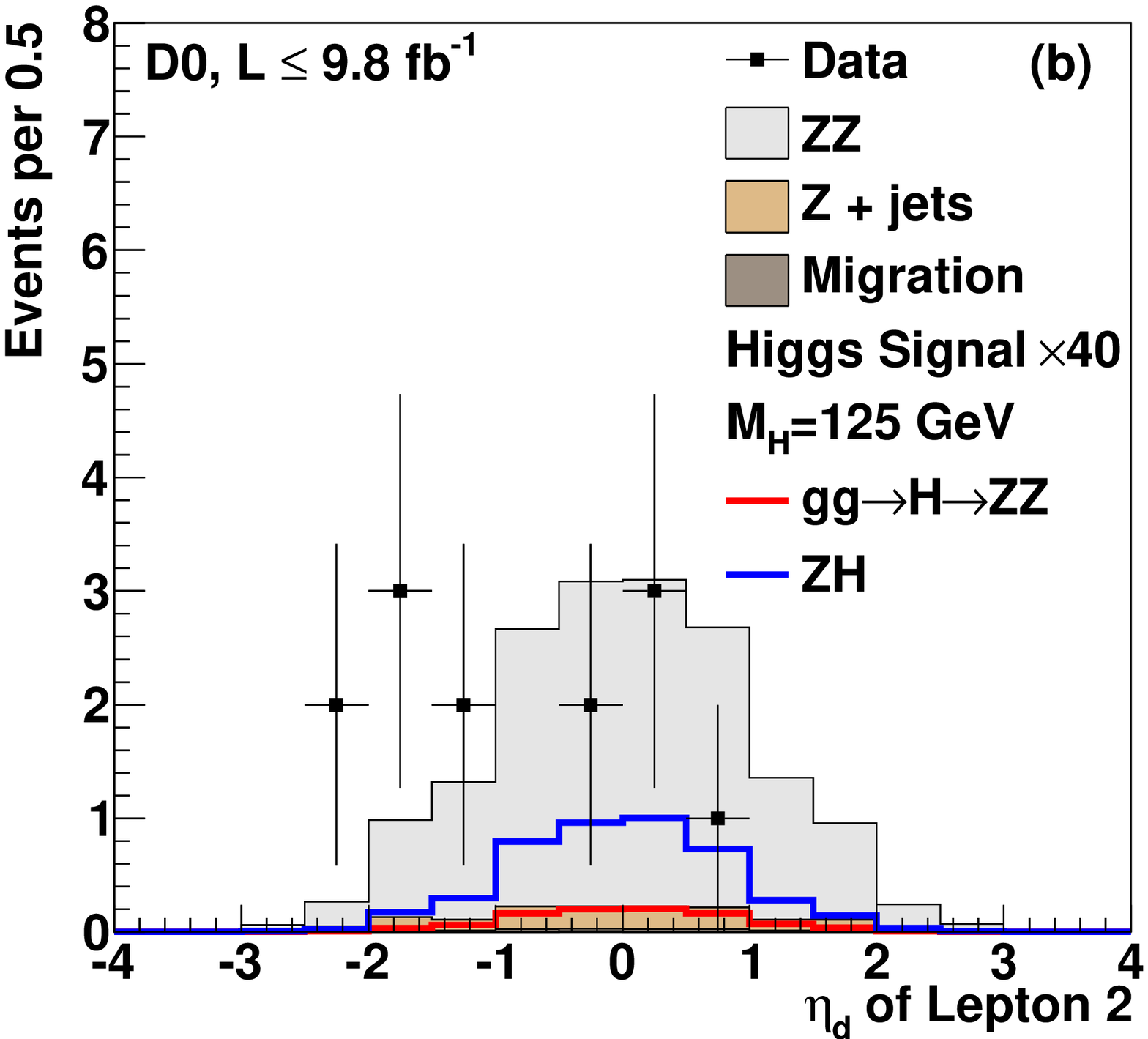}\\
\includegraphics[width=0.40\textwidth]{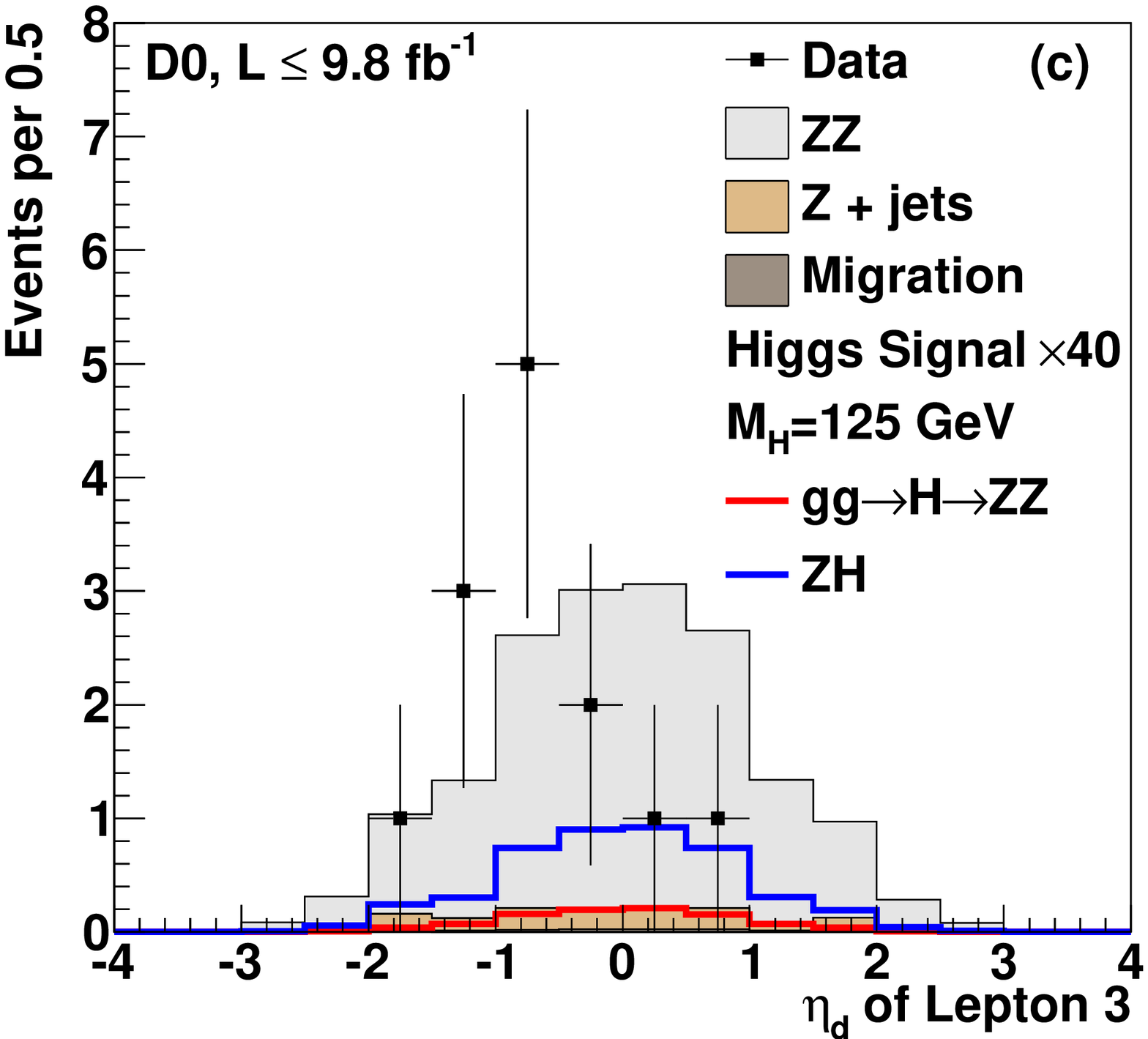}
\includegraphics[width=0.40\textwidth]{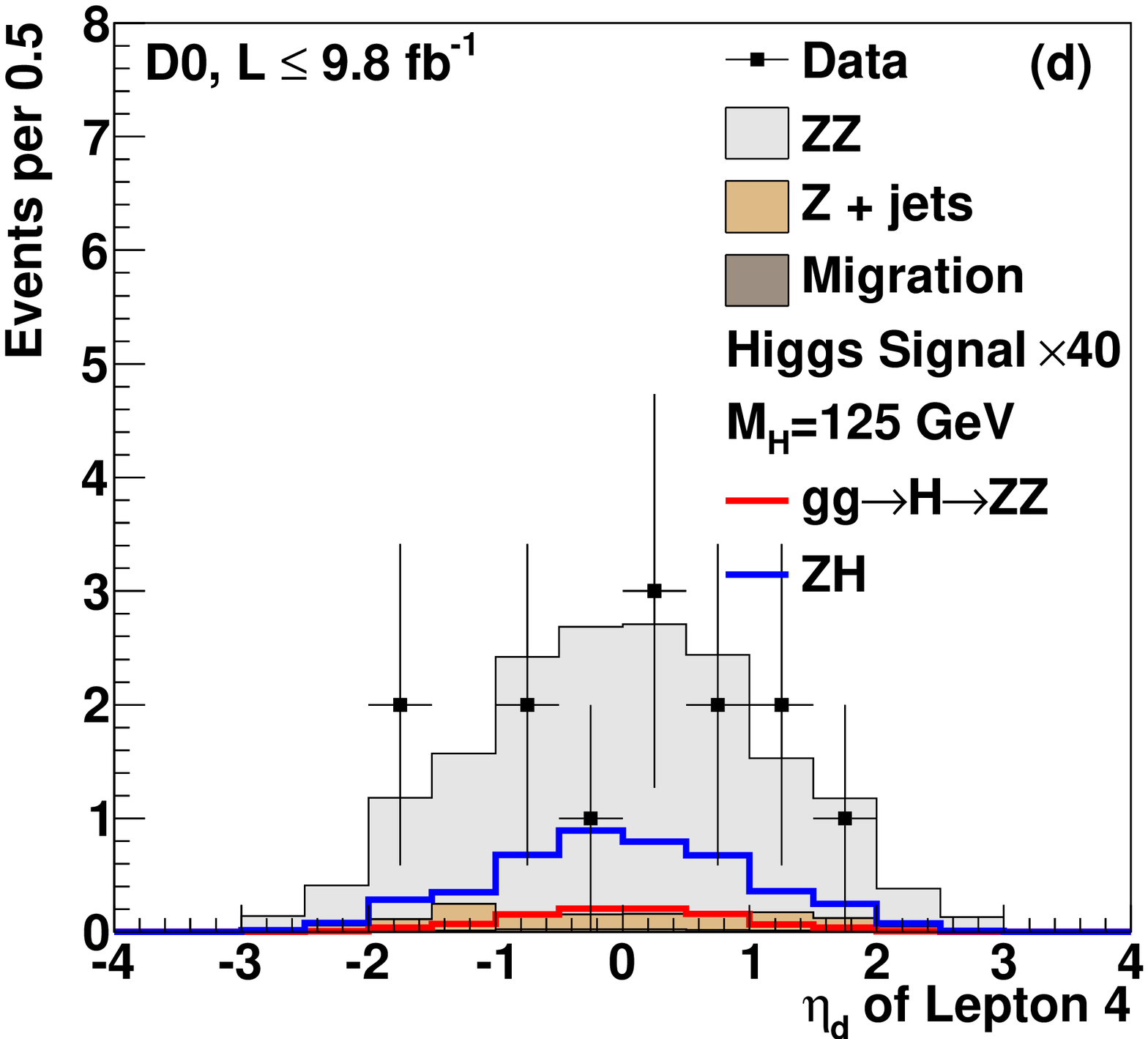}
\caption{ \label{fig:leptoneta} Distributions of $\eta_{d}$ in data, expected signal, and backgrounds for the (a) highest-$p_T$, (b) second-highest-$p_T$, 
(c) third-highest-$p_T$, and (d) lowest-$p_T$ leptons in each event.  
The Higgs boson signal for $M_H$ of 125 GeV is shown scaled by a factor of 40.}
\end{figure*}

\begin{figure*}[!htb]
\centering
\includegraphics[width=0.40\textwidth]{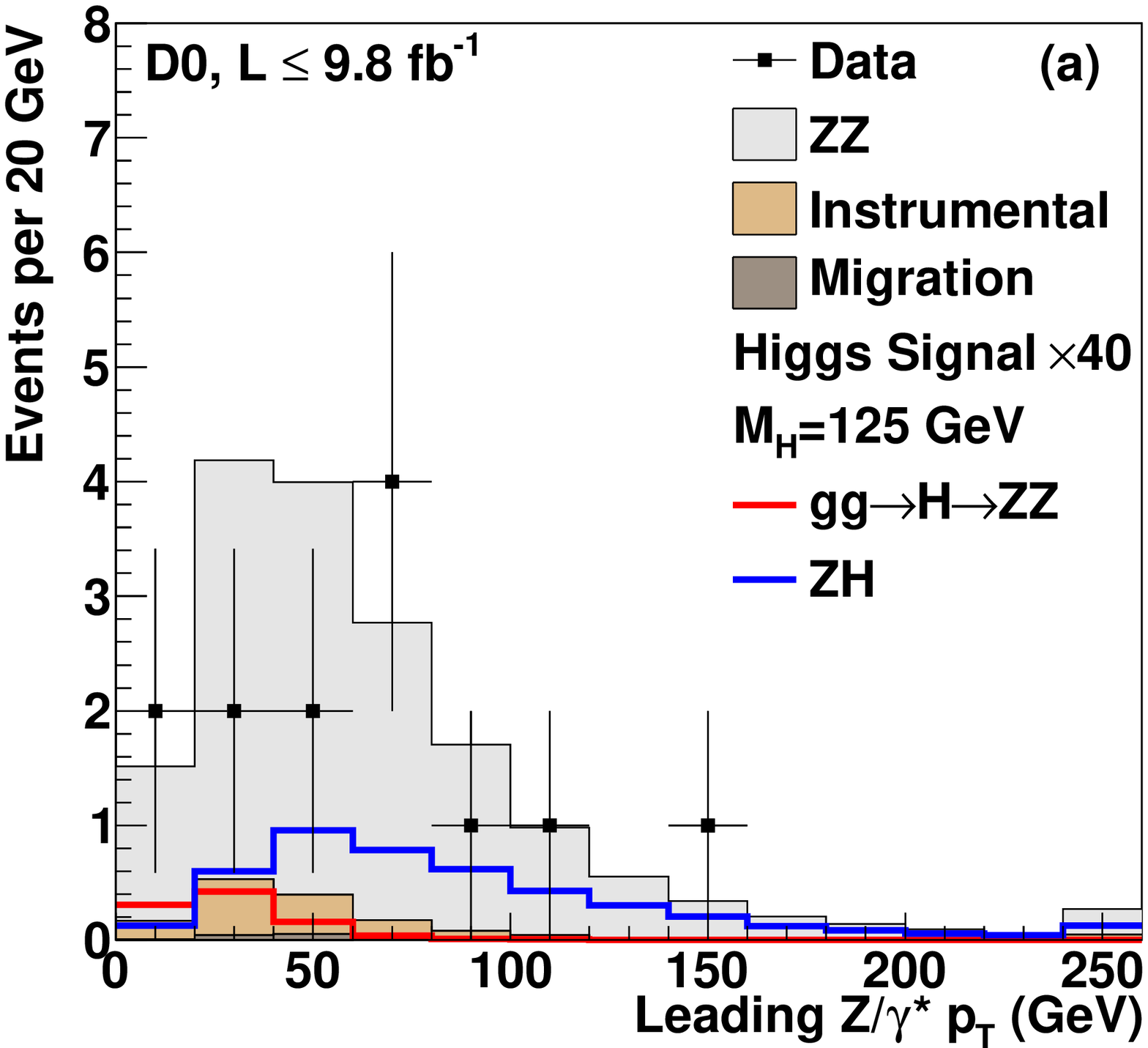}
\includegraphics[width=0.40\textwidth]{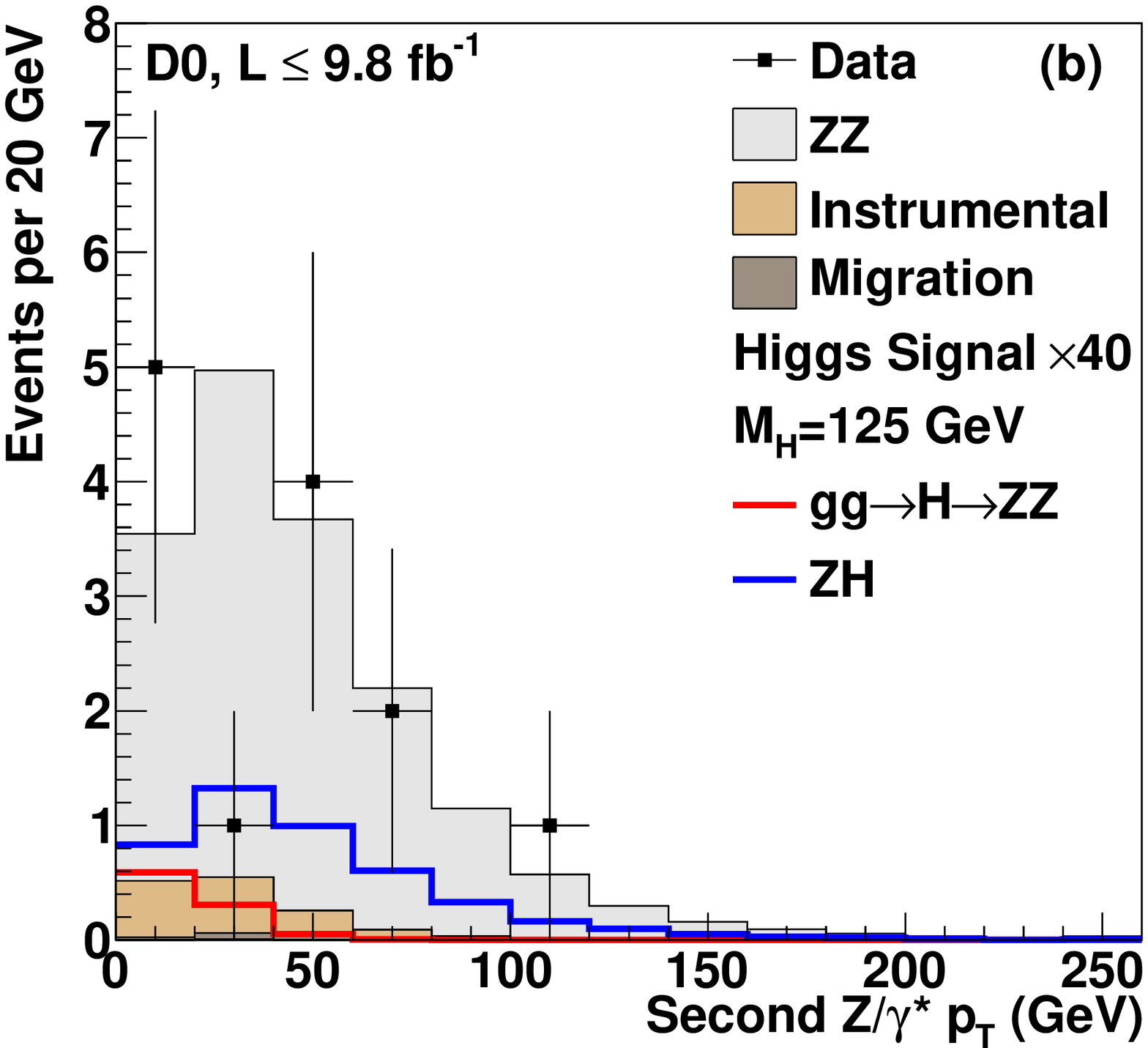}
\caption{ \label{fig:zpts} Distributions of the $Z/\gamma^{*}$ $p_T$ for the (a) leading and (b) second highest-$p_T$ lepton pairings in 
each event.  In the $eeee$ and $\mu \mu \mu \mu$ channels, the combination shown is that with one dilepton mass most consistent with a $Z$ 
mass of 91.2 GeV.
The Higgs boson signal for $M_H$ of 125 GeV is shown scaled by a factor of 40.}
\end{figure*}

\begin{figure*}[!htb]
\centering
\includegraphics[width=0.40\textwidth]{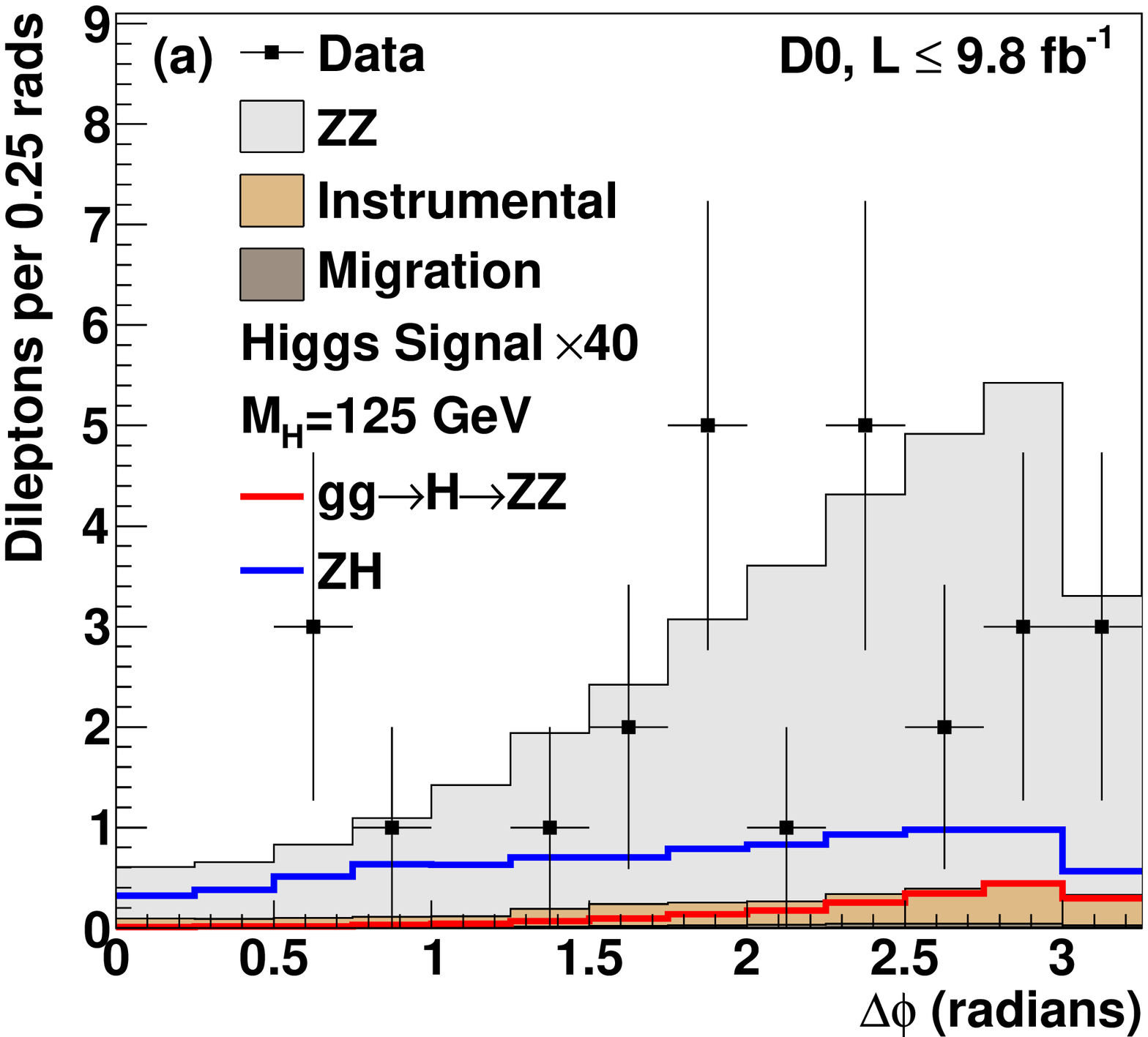}
\includegraphics[width=0.40\textwidth]{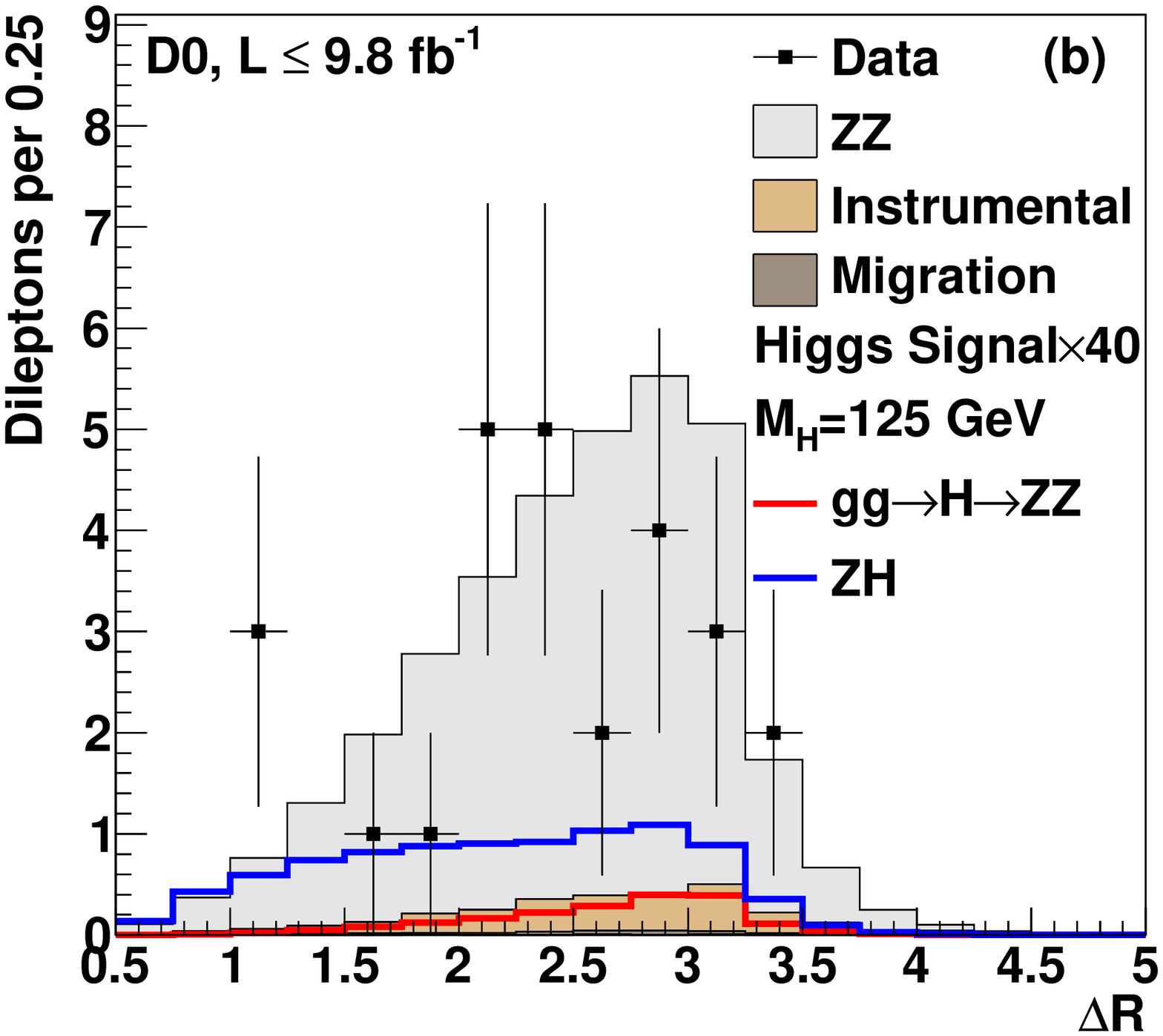}
\caption{ \label{fig:delphiR} Distributions of (a) the opening azimuthal angle, $\Delta \phi$, and (b) the $\Delta R$, between the two leptons of 
a $Z/\gamma^{*}$. 
In the $eeee$ and $\mu \mu \mu \mu$ channels, the combination shown is that with one dilepton mass most consistent with a $Z$ mass of 91.2 
GeV. 
There are two entries per event in both distributions. 
The Higgs boson signal for $M_H$ of 125 GeV is shown scaled by a factor of 40.}
\end{figure*}

\begin{figure*}[!htb]
\centering
\includegraphics[width=0.55\textwidth]{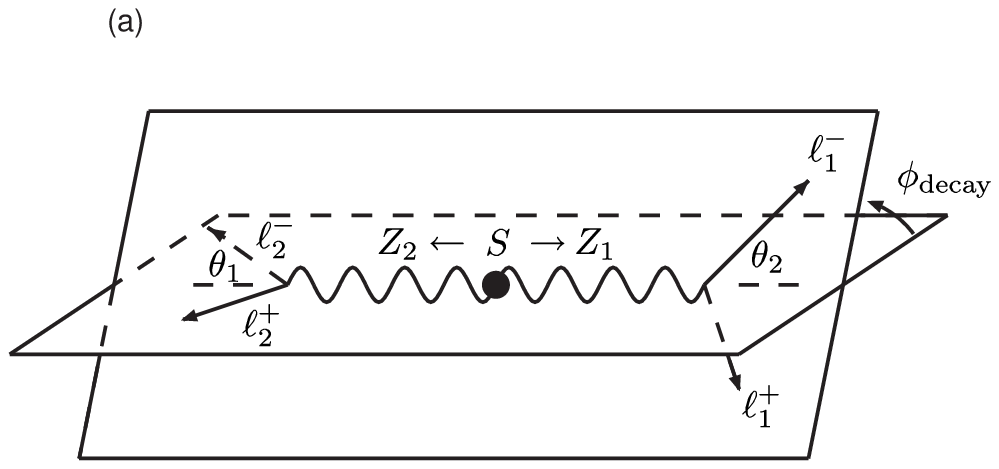}
\includegraphics[width=0.40\textwidth]{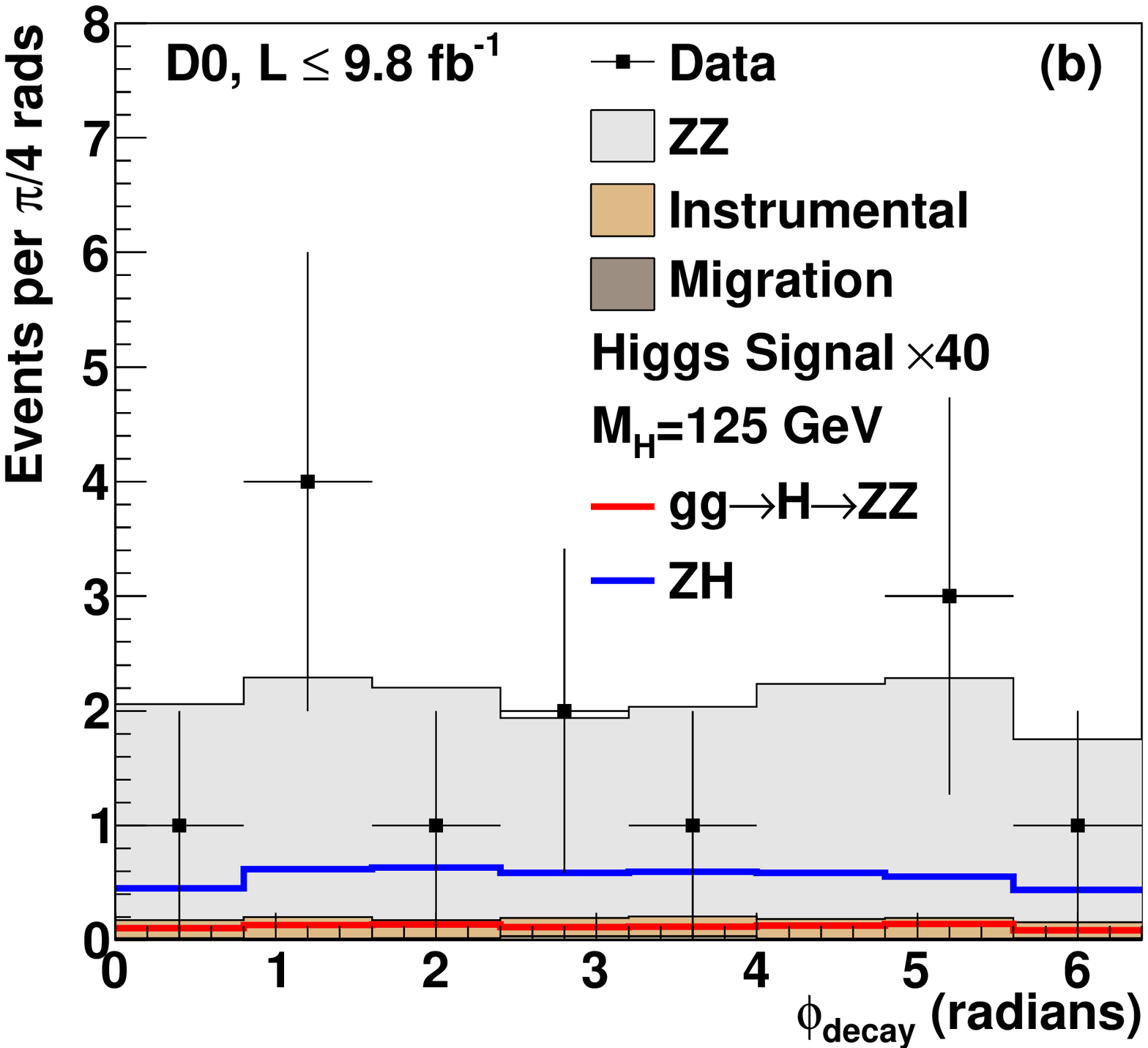}
\caption{ \label{fig:phidecay}  Shown in (a) is definition of $\phi_{\rm decay}$,  
adapted from Ref.~\cite{phidecayref} with permission.
Shown in (b) is the distribution of the azimuthal $\phi_{\rm decay}$ angle.  In the $eeee$ and $\mu \mu \mu \mu$ channels, 
$\phi_{\rm decay}$ is calculated between the combination is most consistent with a $Z$ mass of 91.2 GeV for one of the two dileptons. 
The Higgs boson signal for $M_H$ of 125 GeV is shown scaled by a factor of 40.}
\end{figure*}

Figs.~\ref{fig:dileptonmasszzpt}--\ref{fig:phidecay} show differential distributions of the events used in the $t$-channel $ZZ$ cross section 
measurement and Higgs boson search.  Some of these distributions are kinematic properties of dilepton systems;  in the $ee \mu \mu$ final 
state, the pairings of $ee$ and $\mu \mu$ are always used.  In the $eeee$ and $\mu \mu \mu \mu$ final states, there may be multiple combinations 
passing our selection requirements.  If there are multiple passing combinations, we use the combination that yields 
a dilepton pair with an invariant mass closest to the nominal $Z$ boson mass of 91.2 GeV \cite{pdg}.   Fig.~\ref{fig:dileptonmasszzpt} shows 
the dilepton invariant mass and the $p_T$ of the four-lepton system.  The $p_T$ and $\eta_{d}$ distributions for each lepton in our events
are shown in Fig.~\ref{fig:leptonpt} and~\ref{fig:leptoneta}, respectively.  The $Z/\gamma^{*}$ $p_T$ distributions for the highest-$p_T$ (leading) and second 
lepton pair are in Fig.~\ref{fig:zpts}.
Fig.~\ref{fig:delphiR} shows the distributions of the opening 
angles between the best matched lepton pairs in each event in the azimuthal angle, $\Delta \phi$, and $\Delta R$.  
Fig.~\ref{fig:phidecay} shows the angle $\phi_{\rm decay}$, which is the angle through which the lepton side of one of the $Z/\gamma^{*}$ boson decay planes 
is rotated into the lepton side of the other $Z/\gamma^{*}$ boson decay plane, and measured in the center-of-mass frame of the $Z/\gamma^{*} \mbox{ }Z/\gamma^{*}$
system \cite{phidecayref}.

\begin{thebibliography}{99}
\bibitem{etaref} The pseudorapidity is given by $\eta = -\ln[\tan(\theta/2)]$, where $\theta$ is the polar angle with respect to the proton beam direction.  We define $\eta$ with respect to the $\ppbar$ interaction vertex and $\eta_d$
with respect to the nominal center of the detector.
\bibitem{delRdef} The variable $\Delta R$ between two objects $i$ and $j$ is defined as $\Delta R = \sqrt{(\eta_i - \eta_j)^2 + (\phi_i - \phi_j)^2}$, where $\phi$ is the azimuthal angle.
\bibitem{cit:acoplanarity} The acoplanarity is given by $\phi_1 - \phi_2 - \pi$, where $\phi$ is the azimuthal angle.
\end{thebibliography}
\end{document}